\documentclass{aa}
\usepackage{natbib,appendix,epsfig,graphicx,color}
\usepackage[varg]{txfonts}
\definecolor{Blue}{rgb}{0.06,0.20,0.93}
\bibpunct{(}{)}{;}{a}{}{,}
\def\Ha {H$_\alpha$}
\def\Hbeta {H$_\beta$}
\def\Hgama {H$_\gamma$}
\def\Hdelta {H$_\delta$}

\def\Bralfa {Br$_\alpha$}
\def\Pfgama {Pf$_\gamma$}
\def\Hdelta{H$_\delta$}
\def\SiIIa {Si~{\scriptsize II}~$\lambda$4128}
\def\SiIIb {Si~{\scriptsize II}~$\lambda$4130}
\def\SiIIc {Si~{\scriptsize II}~$\lambda$5041}
\def\SiIId {Si~{\scriptsize II}~$\lambda$5056}
\def\SiIIIa {Si~{\scriptsize III}~$\lambda$4553}
\def\SiIIIb {Si~{\scriptsize III}~$\lambda$4568}
\def\SiIIIc {Si~{\scriptsize III}~$\lambda$4575}
\def\SiIIId {Si~{\scriptsize III}~$\lambda$4813}
\def\SiIIIe {Si~{\scriptsize III}~$\lambda$4820}
\def\SiIIIf {Si~{\scriptsize III}~$\lambda$4829}

\def\SiIVa {Si~{\scriptsize IV}~$\lambda$4089}
\def\SiIVb {Si~{\scriptsize IV}~$\lambda$4116}
\def\SiIVc {Si~{\scriptsize IV}~$\lambda$4212}
\def\SiIVd {Si~{\scriptsize IV}~$\lambda$4631}
\def\SiIVe {Si~{\scriptsize IV}~$\lambda$4950}
\def\SiIVf {Si~{\scriptsize IV}~$\lambda$6668}
\def\SiIVg {Si~{\scriptsize IV}~$\lambda$6701}
\def\SiIVk {Si~{\scriptsize IV}~$\lambda$7048}
\def\SiIVl {Si~{\scriptsize IV}~$\lambda$8957}
\def\heiic {He~{\scriptsize II}~$\lambda$4686}
\def\heiib {He~{\scriptsize II}~$\lambda$4542}
\def\heiia {He~{\scriptsize II}~$\lambda$4200}
\def\heiid {He~{\scriptsize II} $\lambda$6683}
\def\heia {He~{\scriptsize I}~$\lambda$4026}
\def\heib{He~{\scriptsize I}~$\lambda$4388}
\def\heic{He~{\scriptsize I}~$\lambda$4471}
\def\heid{He~{\scriptsize I}~$\lambda$4713}
\def\heie{He~{\scriptsize I}~$\lambda$4922}
\def\heif{He~{\scriptsize I}~$\lambda$6678}

\def\si31{Si~{\scriptsize III}~$\lambda$4553}
\def\si3{Si~{\scriptsize III}~$\lambda\lambda$4553, 4568, 4575}
\def\si4{Si~{\scriptsize IV}~$\lambda\lambda$4089, 4116}
\def\Mdot {$\dot M$\,}

\def\Yhe {$Y_{\rm He}$} 
\def\kms {km~s$^{\rm -1}$\,} 
\def\Rstar {$R_\star$\,}

\def\Teff {$T_{\rm eff}$\,}

\def\vinf {$v_\infty$\,}

\def\vsini {$v \sin i$\,}
 
\def\logg {$\log g$\,}
\def\loggc {$\log g_{\rm c}$\,}
\def\logQ {$\log Q$\,}
\def\Vr {$v_{\rm r}$\,}

\def\logl {$\log L/L_\odot$\,}
\def\vmac {$v_{\rm mac}$\,}
\def\vmic {$v_{\rm mic}$\,}

\def \epsSi {$\epsilon_{\rm Si}$\,}
\def\Msun {$M_\odot$\,}
\def\Rsun {$R_\odot$\,}

\def\Zsun {$Z_\odot$\,}
\def \ao{\frac{1}{\alpha'}}

\def \beq{\begin{equation}}
\def \eeq{\end{equation}}
\def \ben{\begin{enumerate}} 
\def \een{\end{enumerate}} 
\def \beqa{\begin{eqnarray}}
\def \eeqa{\end{eqnarray}}

\def \o{\phantom{0}}
\begin{document}

   \title{The VLT-FLAMES Tarantula Survey}
   \subtitle{XXXII. Low-luminosity late O-type stars - classification, main physical parameters, and silicon  abundances}

\author{N. Markova\inst{1}, J. Puls\inst{2}, P. L. Dufton\inst{3}, D. J. Lennon\inst{4}, 
C. J. Evans\inst{5}, A. de Koter\inst{6}, O. H. Ram\'irez-Agudelo\inst{7}, H. Sana\inst{8} and J. S. Vink\inst{9}}

\offprints{N. Markova,\\ \email{nmarkova@astro.bas.bg}}

\institute{Institute of Astronomy, National Astronomical Observatory,
  Bulgarian Academy of Sciences, P.O. Box 136, 4700 Smolyan, Bulgaria
  \and LMU M\"{u}nchen, Universit\"{a}ts-Sternwarte, Scheinerstrasse
  1, D-81679 M\"unchen, Germany \and ARC, School of Mathematics \&
  Physics, The Queen's University of Belfast, Belfast, BT7 1NN,
  Northern Ireland, UK \and Instituto de Astrof\'isica de Canarias,
  E-38205 La Laguna, Tenerife, Spain \and UK Astronomy Technology
  Centre, Royal Observatory, Blackford Hill, Edinburgh, EH9 3HJ, UK
  \and Astronomical Institute Anton Pannekoek, Amsterdam University,
  Science Park 904, 1098 XH, Amsterdam, The Netherlands \and German  
  Aerospace Center (DLR), Institute for the Protection of Maritime
  Infrastructures, Fischkai 1, D-27572 Bremerhaven, Germany \and Institute
  voor Sterrenkunde, KU Leuven, Celestijnenlaan 200D, 3001, Leuven,
  Belgium \and Armagh Observatory and Planetarium, College Hill,
  Armagh, BT61 9DG, UK } \date{Received 8 November 2019 /Accepted 16 December 2019 } \abstract{Analysis
  of late O-type stars observed in the Large Magellanic Cloud (LMC) by
  the VLT-FLAMES Tarantula Survey (VFTS) revealed a discrepancy
  between the physical properties estimated from model-atmosphere
  analysis and those expected from their morphological
  classifications.  Here we revisit the analysis of 32 of these
  puzzling objects using new hydrogen-helium-silicon {\sc fastwind}
  models and a different fitting approach to re-evaluate their
  physical properties. Our new analysis confirms that these stars
  indeed have properties that are typical of late O-type dwarfs.  We
  also present the first estimates of silicon abundances for O-type
  stars in the 30~Dor clusters NGC~2060 and NGC~2070, with a weighted
  mean abundance for our sample of $\epsilon_{\rm
    Si}$\,$=$\,7.05\,$\pm$\,0.03.  Our values are $\sim$0.20\,dex
  lower than those previously derived for B-type stars in the LMC
  clusters N~11 and NGC~2004 using {\sc tlusty} models. Various
  possibilities (e.g. differences in the analysis methods, effects of
  microturbulence, and real differences between stars in different
  clusters) were considered to account for these results. We also
used our grid of {\sc fastwind} models to reassess the impact of using
the Galactic classification criteria for late O-type stars in the LMC
by scrutinising their sensitivity to different stellar properties. At
the cool edge of the O star regime the \heiic/\heid\ ratio used to
assign luminosity class for Galactic stars can mimic giants or bright
giants in the LMC, even for objects with high gravities
(\logg\,$>$\,4.0\,dex). We argue that this line ratio is not a
reliable luminosity diagnostic for late O-type stars in the LMC, and
that the \SiIVa/\heia\ ratio is more robust for these types.}
 

\titlerunning{Properties of low-luminosity O stars in the VFTS}
\authorrunning{N. Markova et al.}

\maketitle

 %

\section{Introduction}\label{intro}

Bright, massive stars play an important role in the evolution of
galaxies and of the Universe as a whole. The Tarantula nebula (30
Doradus) in the Large Magellanic Cloud (LMC) comprises the most
massive starburst cluster and giant H~{\scriptsize II} region in the
Local Group, and contains the most massive stars known
(\citealt{crowther10}, \citealt{B12}, \citealt{Sch18}). As such,
30~Dor was the target of the VLT-FLAMES Tarantula Survey
\citep[VFTS,][]{evans11}, an unprecedented spectroscopic
survey of 800 OB-type stars investigating their physical properties
and multiplicity.

Spectra of approximately 350 of the VFTS targets were classified as
O-type \citep{walborn14}. Physical properties from atmospheric
analysis (including wind parameters) for the apparently single O-type
dwarfs were presented by \citet{carolina14,carolina17}, with a similar
analysis for the giants and supergiants by \citet[][hereafter RA17]{R17}. 
These studies employed hydrogen--helium models calculated
with the latest version (v10.1) of the {\sc fastwind} (FW) NLTE
line-blanketed, model-atmosphere code \citep{puls05,gonzalez12},
combined with automated fitting using the IACOB--Grid Based Automatic
Tool \citep[{\sc iacob--gbat}, see][]{sergio11} for the former two
studies, and a method employing the {\sc pikaia} Genetic Algorithm
(hereafter GA, \citealt{char95}) for the ostensibly more luminous stars
\citep[see][]{mokiem05}.

This approach proved to be both effective and reliable, but the
resulting parameters were rather unexpected in some cases. In
particular, RA17 reported a group of low-luminosity
(4.20\,$<$\,\logl\,$<$\,5.00) late O-type giants and bright giants
with  effective equatorial surface gravities 
(3.8\,$\le$\logg\,$\le$4.6 dex) that were
greater than expected for their luminosity class\footnote{When stars
  evolve away from the zero-age main sequence their radii increase,
  and therefore their surface gravities decrease. Thus, although still on the
  main sequence, the more evolved O stars (giants and supergiants)
  would be expected to have lower gravities compared to the dwarfs.}.
Some of these were also found to have a helium content that was lower
than expected for such stars in the LMC (see e.g. \citealt{Brott}).
Interestingly, stars with rather high values of \logg\ (4.4 to 4.5)
and peculiarly low helium abundances were also identified among the O
dwarfs from the VFTS \citep{carolina17}.

Three hypotheses have been suggested to explain the puzzling results
for this subset of stars: 
\begin{enumerate}
\item[i)]{Limitations in HHe models for low-luminosity 
late-O stars;} 
\item[ii)] {intricacies in spectral classification, such as for 
example metallicity effects, strong nebular contamination in the Balmer 
and He~I lines, and/or specific difficulties in morphological 
classification at late-O subtypes;} and 
\item [iii)] {undetected binarity or 
contamination in the fibre by another star that is not gravitationally 
bound with the target but impacts on the parameters derived from the 
spectroscopic analysis and available photometry.}
\end{enumerate}

This lack of consistency between morphological classifications and
stellar properties from model atmosphere analyses presents an
interesting challenge for the study of the low-luminosity late-O
stars in the VFTS. As these comprise a reasonable fraction ($\sim$9\%)
of the O stars observed by the survey, we decided to reinvestigate
their properties with an alternative method to those used previously.
This included fits with hydrogen--helium--silicon (HHeSi) FW models and
a classical `by-eye' approach to the model fits. We also reassessed
the classification criteria used by \citet{walborn14} at the cool
end of the O-star regime, focussing on the stellar properties that
most critically influence the relative strengths of the diagnostic
lines.

The layout of the paper is as follows. Section~2 presents the selection
of the sample.  Section~\ref{mod_anal} describes the new model grids and
analysis strategy.  Section~\ref{results} compares the estimated
properties of the sample with published results and presents the first
estimates of silicon abundances in late O-type stars in the LMC.
Section~\ref{spec_class} investigates the predicted behaviour of the
diagnostic lines compared to the classification criteria used for the
sample, and Section~\ref{conclusions} summarises our findings.

\begin{table*}
\begin{center}
\caption[]{Stellar sample with classifications and additional morphological 
comments from \citet{walborn14}.}
\label{sample}
\tabcolsep2.10mm
\begin{tabular}{llcll}
\hline\hline
VFTS & Classification & Rating$^{(a)}$ & Mult?$\,^{(b)}$ & Comments \\
\hline
070  &O9.7 II                    & BBB  &SB? & Si {\scriptsize IV} very weak\\
077  &O9.5: IIIn                 & BBB  &    & Si {\scriptsize IV} very weak\\
080  &O9.7 II-III((n))           & BBB  &    & Si {\scriptsize IV} weak\\
091  &O9.5 IIIn                  & AAA  &    & \\%
104  &O9.7 II-III((n))           & BBB  &    & Si {\scriptsize IV} very weak\\
113  &O9.7 II or B0 IV           & BBB  &SBs & Si {\scriptsize IV} very weak\\
128  &O9.5 III:((n))             & BBB  &    & Si {\scriptsize IV} weak\\
131  &O9.7                       & BBB  &    &\\
141  &O9.5 II-III((n))           & BBB  &SB? & Si {\scriptsize IV} very weak\\
188  &O9.7: III                  & BBB  &SB? & Si {\scriptsize IV} very weak\\
192  &O9.7 II or B0 IV           & BBB  &    & Si {\scriptsize IV} weak\\
205  &O9.7 II((n)) or B0 IV((n)) & BBB  &    & Si {\scriptsize IV}, C {\scriptsize III} very weak\\
207  &O9.7 II((n))               & BBB  &    & Si {\scriptsize IV}, C {\scriptsize III} very weak\\
210  &O9.7 II-III((n))           & AAA  &    & \\
226  &O9.7 III                   & BBB  &    & Si {\scriptsize IV} weak; runaway\\
328  &O9.5 III(n)                & BBB  &SB? & Si {\scriptsize IV} weak; runaway\\
346  &O9.7 III                   & BBB  &    & Si {\scriptsize IV} very weak\\
370  &O9.7 III                   & BBB  &SB? & Si {\scriptsize IV} weak; runaway\\
412  &O9.7                       & BBB  &    &\\
444  &O9.7                       & BBB  &SB? &\\
495  &O9.7 II-IIIn               & BBB  &    & Si {\scriptsize IV} very weak\\
528  &O9.7(n)                    & BBB  &    & Si {\scriptsize IV} very weak\\
569  &O9.2 III:                  & AAA  &    & Si {\scriptsize IV} weak\\
571  &O9.5 II-III(n)             & BBB  &SB? & Si {\scriptsize IV} weak\\
574  &O9.5 IIIn                  & AAA  &SB2? & \\
607  &O9.7 III                   & BBB  &SB? & Si {\scriptsize IV} very weak\\
615  &O9.5 IIInn                 & AAA  &VM3 & \\
620  &O9.7 III(n)                & AAA  &SB? & Si {\scriptsize IV} weak\\
622  &O9.7 III                   & BBB  &SB? & Si {\scriptsize IV} weak\\
753  &O9.7 II-III                & AAA  &    & \\
787  &O9.7 III                   & BBB  &SB  & Si {\scriptsize IV} weak\\
843  &O9.5 IIIn                  & AAA  &    & \\
\hline
\end{tabular}
\end{center}
\small
{\bf Notes.} (a) AAA and BBB indicate high-/low-rated classifications, respectively.
(b) Comments from \citet{walborn14} on possible multiplicity: `SB?' -- stellar absorption 
displaced from nebular emission lines but no radial-velocity variation measured; 
`SBs' -- small amplitude (10--20\,\kms) radial-velocity variations; 
`SB2?' -- possible double-lined binary system; `VM3' -- visual multiple consisting 
of three components.
\end{table*}
\normalsize

\section{Observational sample}\label{sample_selection}

The sample from RA17 comprised 72 presumably single O-type giants and
supergiants, complemented with 31 stars classified as O-type but
without luminosity classifications \citep[see][]{walborn14}. For a
remarkable 58\% of the former, the estimated equatorial surface
gravities do not match values that would be expected from their
luminosity classifications.  

The RA17 sample was assembled by including targets with no significant
radial velocity (\Vr) variations, or where significant but only small
shifts ($\Delta$\Vr\,$<$\,20\,\kms) were seen from the VFTS
spectroscopy \citep{sana13}. Several of these objects were later
identified as binaries by further monitoring by \citet{almeida17} and
were omitted from our sample\footnote{\citet{almeida17} found
  $\Delta$\Vr\,$=$\,28\,\kms for VFTS\,113 but were unable to estimate
  a reliable period so we retained it in our sample as an apparently
  single star.}.

Of course, some of our objects will probably be identified in the
future as binaries. \citet{sana13} estimated an intrinsic binary
fraction of $\sim$50\% for the O-type stars from the VFTS. Therefore, in
the sample studied by RA17, at least 32 binaries (with periods less
than about 10\,yr) may be expected, with only six idenfied so far by
\citet{almeida17}. The VFTS spectra might also include contributions
from more than one star (not necessarily physically bound), although
unless the two components have similar effective temperatures and
luminosities, the analysis presented here should reveal
peculiarities arising from binarity or composite spectra.

\medskip

Discarding the most recently discovered binaries and objects with 
poor-quality fits (see Tables C.4 and C.5 in RA17, respectively),  
32 objects remain from RA17 with peculiarly high surface
gravities, which form the observational sample for our study. 
Their classifications from \citet{walborn14} are listed in
Table~\ref{sample}, with all but one classified as O9.5-O9.7.  The
table also includes additional morphological comments of note as follows:
\begin{itemize}
\item{{\it Rating:} The `AAA' notation was used by
    \citeauthor{walborn14} for spectra with sufficiently high quality
    for morphological analysis. A `BBB' rating indicated objects with
    low-rated classifications due to for example low signal-to-noise (S/N)
    ratios, strong nebular contamination in the Balmer and He~I lines,
    or suspected binarity.}
\item{{\it Additional comments regarding multiplicity:}
    \citeauthor{walborn14} also noted morphological features that may
    indicate evidence for binary companions that were otherwise
    undetected from the \Vr\ analyses.  As indicated in
    Table~\ref{sample}, these primarily comprised stars where a clear
    \Vr\ offset was seen between the stellar absorption and nebular
    emission lines (`SB?').}
\item{{\it Si weak stars:} Many of the sample spectra also displayed a
    discrepancy between the \heiic/\heid\ and the \SiIVa/\heia\ line
    ratios used to assign luminosity classes (cf. Galactic standards) for
    late O-types.  These were flagged as `Si~{\scriptsize IV} weak' by
    \citeauthor{walborn14}}
\end{itemize}

The majority of the sample are members of the two main clusters in
30~Dor (NGC~2070 and NGC~2060)\footnote{\citet{sana13} defined radii
  of 2\farcm4 for membership of these two clusters and we adopted the
  same definition here.}, with only nine located in the nearby field.
As seen from Table~\ref{sample}, 24 (75\%) of our targets were BBB
rated, highlighting that many of the stars with unexpected parameters
are those with lower quality data. It is also notable that 23 (72\%)
were classified as Si weak (all but two of which were also rated as
BBB), and 14 (44\%) have some indication of possible multiplicity or
composite nature.

\medskip

To improve the S/N ratios and to simplify the analysis, RA17 employed
spectra where the multi-epoch observations for each target were
combined into a single, normalised spectrum.  For consistency we use
the same spectra here.

\section{Model atmosphere analysis }\label{mod_anal}

Helium is the traditional temperature indicator in atmospheric
analysis of O-type stars. Using HHe models was effective for the
majority of the O stars from the VFTS, but there were cases where it
did not provide robust results, such as weak or absent helium lines or
spectra with strong nebular contamination.  Alternative diagnostics
such as nitrogen (early to mid-O stars) and silicon (O9.2-O9.7 stars)
were therefore used to provide additional constraints
\citep[][RA17]{carolina14,carolina17}.

As our sample consists only of late O-type stars, we employed HHeSi
models calculated with FW (v10.1) to analyse the spectra. For an
initial assessment of the basic stellar parameters, namely effective
temperature (\Teff), the logarithm of the effective surface gravity 
(\logg), microturbulence (\vmic), helium and silicon content 
with respect to particle number (\Yhe = $N({\rm He)}/N({\rm H})$ and 
\epsSi = $\log[N{(\rm Si)}/N({\rm H})] + 12$, respectively),
and wind-strength parameter $Q$ = \Mdot/(\vinf*\Rstar)$^{1.5}$ (see 
\citealt{puls96}), we used a newly constructed grid of FW models with 
parameters typical for late-O and early-to-mid B stars in the LMC.  
For all but microturbulent velocity (see Sect.~\ref{para_determinations}), 
these estimates were additionally fine-tuned by calculating a 
higher-resolution grid around the initial parameters, with tailored values 
of \Teff, \logg, \Yhe, \epsSi, and \Mdot. 

\subsection{The new FASTWIND grid}\label{grid}

Non-local thermodynamic equilibrium (NLTE), line-blanketed, spherically 
symmetric models with stellar winds
are generally used to analyse O-type stars (e.g. \citealt{crowther02}
\citealt{massey04, massey05}; \citealt{mokiem05}). In contrast,
following tests of results from static versus  wind models by
\citet{dufton05}, NLTE, line-blanketed, plane-parallel models ({\sc
  tlusty/synspec}) are often used to investigate B-type stars
\citep[e.g.][]{hunter07,trundle07,mcevoy15,dufton18}. This can lead to
difficulties in comparing results from the two approaches, as
discussed later in Sect.~\ref{Si}.

The new grid was calculated to provide a tool to explore the critical
regime of late O- and early B-type stars in the LMC, in a
statistically significant and internally consistent manner, and taking
into account the effects of stellar winds in each object. The grid
comprises $\sim$6\,000 atmosphere models with more than 20\,000
synthetic spectra, computed using seven input parameters: \Teff,
\logg, \Yhe, \epsSi, microturbulent velocity (\vmic), the wind
velocity field exponent $\beta$, and $Q$. A global metallicity (though
specifying the helium and silicon content explicitly) of $Z$\,$=$\,0.5~\Zsun\
was adopted (see e.g. \citealt{mokiem07})
\footnote{Reference solar abundances were taken from
  \citet{asplund09}.}.

The parameter space of the grid is summarised in Table~\ref{FWgrid}.
Detailed information on how each of the listed parameters was
specified in the calculations will be presented in a forthcoming paper
(Petrov et al. in prep.). Here we briefly note specific points that
are pertinent to the current analysis:
\begin{itemize}
\item[i)]{To account for potential problems 
    with the Si model atom,
    \citep[previously noted by e.g.][]{urbaneja05}, an improved
    Si~{\scriptsize IV} atomic model (from \citealt{pauldrach94}) was
    implemented in the code, with other Si ions taken from
    \citet{MP08}.}    
\item[ii)] {\citet{paco11} demonstrated that profile fits for stars with
    thin winds (such as O dwarfs in the LMC) do not require the
    inclusion of clumping, and so all models were computed under the
    assumption of homogeneous outflows. For stars with stronger winds,
    this might lead to overestimates of \Mdot.}
\item[iii)] {In model atmosphere computations, the microturbulent
    velocity affects the NLTE occupation numbers (and thus also the
    atmospheric structure), and the formal integral calculations
    (emergent profiles). To compute the grid, we assumed a
    depth-independent, microturbulent velocity of 10\,\kms\ for the
    NLTE part\footnote{In the parameter range of (non-supergiant)
      O-type stars, a \vmic-value of the order of 10~\kms seems to be
      consistent with a variety of investigations (see e.g
      \citealt{repo}, \citealt{gonzalez12}, \citealt{markova18},
      \citealt{massey13}).}, and of 5, 7, 10, and
    15~\kms for the formal integral, to cover the range of estimates
    derived for B stars in the LMC (see Fig.~\ref{fig5} and references
    therein).}
\item[iv)] {Potential (micro-)turbulence pressure terms are not accounted
    for in the calculation of the photospheric structure. This might
    lead to underestimated surface gravities (by up to 0.1-0.15\,dex,
    see e.g. \citealt{markova18}).}
\end{itemize}

\begin{table}
\begin{center}
\caption[]{Parameter space covered by the new LMC HHeSi model grid.}\label{FWgrid}
\tabcolsep1.20mm
\begin{tabular}{ll}
\hline\hline
\multicolumn{1}{l}{Parameter} &\multicolumn{1}{l}{Ranges and/or specific values}\\
\hline
\Teff  & 36 to 16\,kK, stepsize of 2\,kK\\
\logg  & 4.2 to 2.0, stepsize of 0.2\,dex\\
\Yhe   & 0.1, 0.15, 0.20\\
\epsSi & 7.0, 7.2, 7.4\\
\vmic  &  10\,\kms\ (atmospheric structure)\\
       & 5, 7, 10 and 15\,\kms\ (formal integral)\\
\logQ$^{(a)}$  & $-$14, $-$13.50, $-$13.15, $-$12.80, $-$12.45\\
$\beta^{(b)}$  & 0.9, 1.0, 1.1, 1.2, 1.3, 1.4, 1.5, 1.6, 2.0, 3.0\\
\hline
\end{tabular}
\end{center}
\small{\bf Notes.} (a) Q is calculated following \citet{puls96} 
with \Mdot\ in \Msun/yr, \Rstar in \Rsun and \vinf in \kms.
(b) Different values used for different temperature regimes.
\normalsize
\end{table}

\subsection{Spectral line diagnostics and fitting techniques}
\label{line_diagnostics}

The FW grid includes 33 lines from H, He, and Si that could be used in a
quantitative analysis (Table~\ref{diagnostic_lines}). However, given
strong nebular contamination of many of the spectra, weak line
transitions, and the limited wavelength coverage of the data, only 17
were available in our analysis (asterisks in Table~\ref{diagnostic_lines}).

\begin{table}
\begin{center}
  \caption[]{Diagnostic spectral lines available in the FW model grid
    for late O and early-to-mid B stars in the LMC.}
\label{diagnostic_lines}
\tabcolsep1.20mm
\begin{tabular}{lll}
\hline
\hline
\multicolumn{1}{l}{Balmer series}
&\multicolumn{1}{l}{Helium}
&\multicolumn{1}{l}{Silicon}\\
\hline
\Ha     &\heia* & \SiIIa\\
\Hbeta  & \heib*  & \SiIIb\\
\Hgama*  &\heic*  & \SiIIc\\
\Hdelta* &\heid*  & \SiIId\\
        &\heie*  & \SiIIIa*\\
        & \heif*  & \SiIIIb*\\
        &\heiia*  & \SiIIIc*\\       
        &\heiib*  & \SiIIId\\     
        &\heiic*  &  \SiIIIe\\     
        &\heiid  & \SiIIIf\\     
        &   & \SiIVa*\\
        &  & \SiIVb*\\
        &  & \SiIVc*\\       
        &  & \SiIVd\\     
        &  & \SiIVe\\     
        &  & \SiIVf\\    
        &  & \SiIVg\\     
        &  & \SiIVk\\    
        &  & \SiIVl\\         
\hline
\end{tabular}
\end{center}
{\small {\bf Notes.} ($\ast$) Diagnostic lines used in the present analysis.}
\end{table}

Spectral line fitting is an optimisation problem where one tries to
maximise the correspondence between observed and synthetic profiles by
minimising the differences between them. There are two ways to deal
with this issue: one may either apply automated fitting techniques such as
those used by \citet{carolina14, carolina17} and RA17, or use the
classical `fit-by-eye' approach where the quality of the fit is
estimated by visual inspection of the strategic lines. We opted for
the latter approach as our goal was an independent study of the selected
objects compared to the results from RA17.

\subsection{Determination of physical parameters}\label{para_determinations}

The stellar and wind properties of our targets were estimated as
follows:

\smallskip 

-- {\textit{Effective temperature} (\Teff): This} was estimated from 
the silicon (Si~{\scriptsize III}/{\scriptsize IV}) ionisation balance, 
with helium (He~{\scriptsize I}/{\scriptsize II}) used as a secondary
check. The estimates from the two elements generally agreed to within
$\pm$1000\,K, which we adopted as the typical uncertainty, except for
the fast rotators (\vsini\,$>$\,150\,\kms) where a larger error of
$\pm$1500\,K was estimated. An error of $\pm$2000\,K was adopted for
three stars where we failed to obtain good fits to \heiic\ in parallel 
with the other lines (see below).

\smallskip

\begin{figure}
{\includegraphics[width=8.5cm,height=5.5cm]{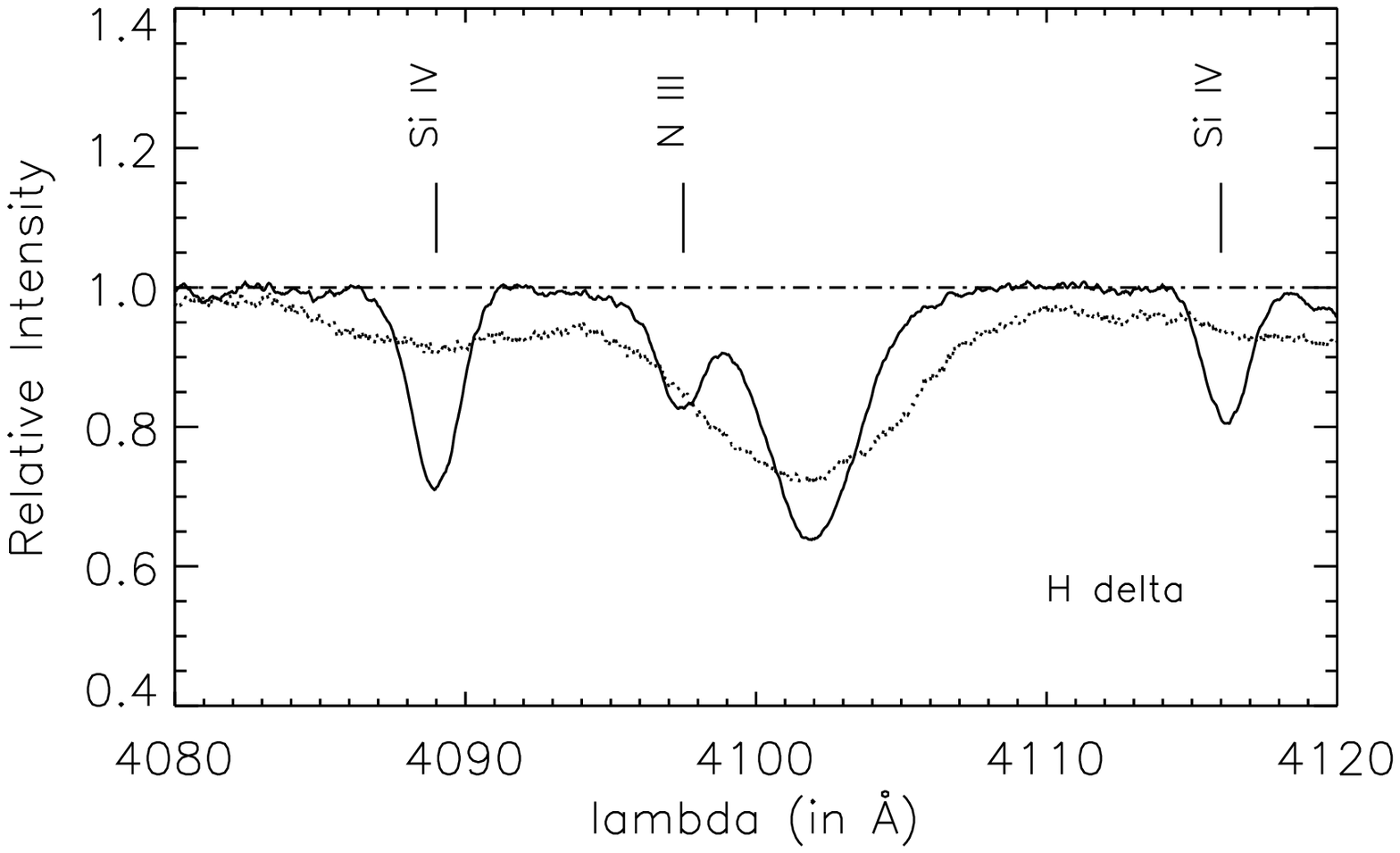}}\\
{\includegraphics[width=8.5cm,height=5.5cm]{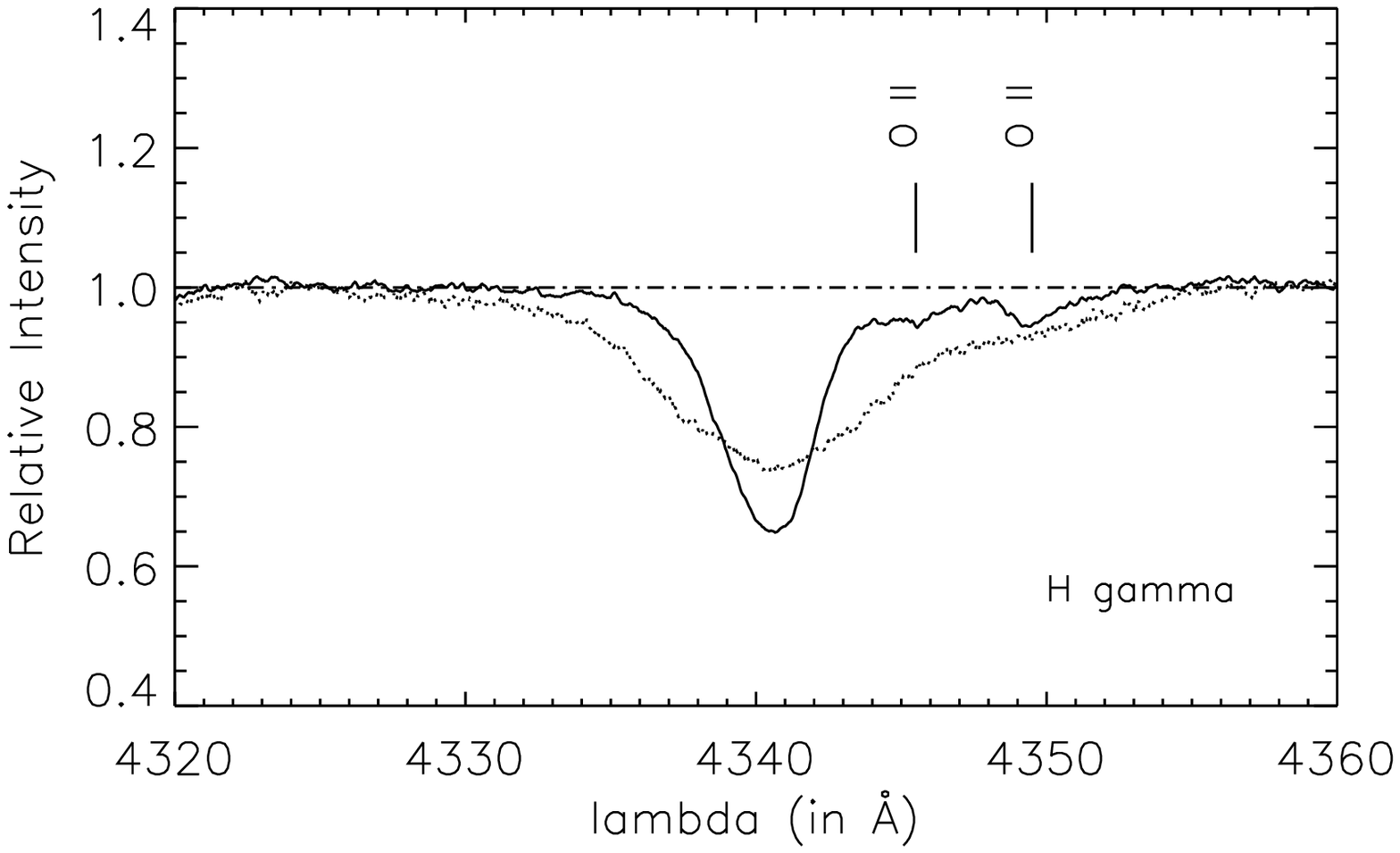}}
\caption{Example \Hdelta\ and \Hgama\ lines that highlight the
  contributions from N~{\scriptsize III} and O~{\scriptsize II} in the
  blueward and redward wings, respectively. To clearly demonstrate the
  potential contamination, the displayed profiles are taken from
  spectra of two Galactic stars, with \vsini\,$=$\,70\,\kms (solid)
  and \vsini\,$=$\,310\,\kms (dotted); less severe but still
problematic contributions are also present at the metallicity of the LMC.}
\label{fig1}
\end{figure}

-- {\it Equatorial surface gravity (\logg):} The best gravity 
diagnostics available in the optical spectra of hot massive stars are
the wings of \Hdelta\ and \Hgama. These appeared to be free of nebular
contamination in our sample, but in the late O-type regime they are
also contaminated by absorption from weak metal lines.  The effects of
these metal lines need to be carefully accounted for (particularly at
higher \vsini) to obtain robust results (see Fig.~\ref{fig1}). With
this in mind, we checked each spectrum and found that, even at fast
rotation rates, the redward \Hdelta\ wing and the blueward \Hgama\
wing are unaffected by metal-line absorption. We therefore gave more
weight to these regions when estimating \logg. The typical error on
\logg\ ranges from $\pm$0.1\,dex for objects with
\vsini\,$\leq$\,150\,\kms to $\pm$0.15\,dex for those with
\vsini\,$>$\,150\,\kms.

\smallskip

-- {\it Stellar radius (\Rstar):} To estimate the Newtonian gravity,
\loggc\ (i.e. the gravity corrected for centrifugal acceleration; see,
e.g.  \citealt{repo}),  knowledge of the stellar radius is required. In
principle, our approach allows this parameter to be estimated for each
star, provided the de-reddened absolute magnitude in the $V$-band is
known or can be calculated using corresponding photometry and
distances from the literature.  The distance and $V$-band magnitudes
for our targets are known, but their visual extinction is uncertain in
many cases \citep{MA14}. Therefore, to calculate the true gravity, we
adopted the \Rstar\ estimates from RA17 which were based on 
de-reddened absolute  magnitudes in the K-band\footnote{These 
magnitudes have been calculated using observed K-band magnitudes 
from \citet{R12}, a distance modulus to the Tarantula nebula of 
18.5 mag from \citet{evans11} and an average K-band extinction of 0.21 
mag from \citet{MA14}}. For our sample, these estimates range from $\sim$4 to
$\sim$10~\Rsun, in perfect agreement with the values expected for 
low-luminosity, unevolved O stars in the LMC (see \citealt{Brott}).

\smallskip

-- {\it Wind parameters:}
The main diagnostics of \Mdot\ for O-type spectra in the optical are
\Ha\ and \heiic, where the former can also be used to constrain
$\beta$. The \Ha\ profiles were strongly contaminated by nebular
emission for many of our sample stars. Therefore, for the purposes of the
current analysis, we fixed $\beta$\,$=$\,0.9 (the prototypical value
for low-luminosity O-stars) and estimated $Q$ from the best fit to
\heiic\footnote{As pointed out by \citet{puls96}, the wind-strength
  parameter $Q$ is derived directly from the spectroscopic analysis,
  while the actual \Mdot\, can only be determined when \Rstar\ and
  \vinf\ are known.  Since \vinf\, is unknown for our targets, we
  consider $Q$ in the present analysis.}. The typical uncertainty on
log\,$Q$, estimated from the quality of the fit to \heiic\ (the small
influence of $\Delta$\Teff was neglected) ranges from 0.2 to 0.3~dex.

The estimated log\,$Q$ values are listed in Col.~8 of Table~\ref{para},
with only upper limits possible for 11 stars\footnote{This is mainly
  because \heiic\ (and also \Ha) becomes insensitive to changes in
  $Q$ below a certain value of \Mdot.}. We caution the
reader that due to the dependence of \Mdot\ on $\beta$, these
estimates are only valid for this specific value of $\beta$.
Three stars 
are flagged with an asterisk in Col.~1 of Table~\ref{para} as we were
unable to obtain a good fit to \heiic\ in parallel with the other
diagnostic lines in the spectrum: the \heiic\ lines appear  asymmetric, weaker than
predicted by the best models, and blueshifted compared to
the corresponding \Vr. Interestingly, similar problems were reported
by RA17 for their larger sample. The reasons for this are unclear,
but might involve binarity or additional line emission of unknown
origin. For these three stars we derived two estimates, representing
upper and lower limits to log\,$Q$, the mean of which is given in
Table~\ref{para}.

\smallskip

{\small
\begin{center}
\begin{table*}
  \caption[]{Estimated physical parameters of our sample of LMC late O-type stars, grouped according 
    to their cluster membership.  In addition to standard abbreviations, \loggc\ 
    is the centrifugally corrected surface gravity. Estimates of log\,$Q$ were derived assuming 
    unclumped winds and $\beta = 0.9$. For all objects, $Y_{\rm He} = 0.1
    \pm 0.02$ was found. Uncertainties in our \vsini estimates range from 10 
    to $\sim$20\%, with higher values typical for more rapid rotators.\label{para}}
\tabcolsep0.95mm
\begin{tabular}{llcccccrccc}
\hline\hline
\multicolumn{1}{l}{VFTS} &\multicolumn{1}{l}{Classification} &\multicolumn{1}{c}{Location } &\multicolumn{1}{c}{\Teff}
&\multicolumn{1}{c}{\logg} &\multicolumn{1}{c}{\loggc$^{(1)}$} &\multicolumn{1}{c}{\Rstar$^{(2)}$}
&\multicolumn{1}{c}{log\,$Q$} &\multicolumn{1}{c}{\vsini$^{(3)}$} &\multicolumn{1}{c}{\Vr} &\multicolumn{1}{c}{\epsSi} \\
\multicolumn{1}{l}{} &\multicolumn{1}{c}{} &\multicolumn{1}{c}{} &\multicolumn{1}{c}{(kK)} &\multicolumn{1}{c}{(cm\,s$^{-2}$)}
&\multicolumn{1}{c}{(cm\,s$^{-2}$)}
&\multicolumn{1}{c}{(\Rsun)}
&\multicolumn{1}{c}{}
&\multicolumn{1}{c}{(km/s)}
&\multicolumn{1}{c}{(km/s)}
&\multicolumn{1}{c}{} \\
\hline
346  &O9.7 III   &NGC~2070   &31.4\,$\pm$\,1.0 &4.15\,$\pm$\,0.10 &4.15 &6.44 &$-$12.75\,$\pm$\,0.20  &\o75 &265 &7.00\,$\pm$\,0.20\\
412  &O9.7       &NGC~2070   &31.0\,$\pm$\,1.0 &4.00\,$\pm$\,0.10 &4.00 &5.40 &$-$12.68\,$\pm$\,0.2   &{\it \o80} &270 &7.10\,$\pm$\,0.10\\
444  &O9.7       &NGC~2070   &31.0\,$\pm$\,1.0 &4.15\,$\pm$\,0.10 &4.16 &7.22 &$-$12.57\,$\pm$\,0.2   &100      &265 &7.10\,$\pm$\,0.10\\
528  &O9.7(n)    &NGC~2070   &30.8\,$\pm$\,1.0 &4.15\,$\pm$\,0.10 &4.16 &7.65 &$-$12.93\,$\pm$\,0.2   &150      &270 &7.00\,$\pm$\,0.20\\
569  & O9.2 III:  &NGC~2070   &32.6\,$\pm$\,1.0 &3.90\,$\pm$\,0.10 &3.90 &7.49 &$-$12.74\,$\pm$\,0.20 & {\it \o35}      &255 &7.00\,$\pm$\,0.10\\
571*  &O9.5 II-III(n)&NGC~2070&32.8\,$\pm$\,2.0 &4.25\,$\pm$\,0.10 &4.26 &5.46 &<$-$13.22         &148      &260 &6.90\,$\pm$\,0.20\\
607  &O9.7 III   &NGC~2070   &32.0\,$\pm$\,1.0 &4.10\,$\pm$\,0.10 &4.10 &5.97  &$-$12.64\,$\pm$\,0.20 & \o60      &265 &7.00\,$\pm$\,0.15\\
615  &O9.5 IIInn &NGC~2070   &32.0\,$\pm$\,1.5 &4.00\,$\pm$\,0.15 &4.08 &10.30\o &$-$12.87\,$\pm$\,0.20 &372      &270 &7.00\,$\pm$\,0.20\\
620  &O9.7 III(n)&NGC~2070   &32.0\,$\pm$\,1.5 &4.10\,$\pm$\,0.15 &4.15 &4.82  &$-$13.15\,$\pm$\,0.20 &208      &275 &7.00\,$\pm$\,0.20\\
622  &O9.7 III   &NGC~2070   &32.0\,$\pm$\,1.0 &4.20\,$\pm$\,0.10 &4.21 &4.64  &$-$12.86\,$\pm$\,0.20 &\o84 &265 &7.00\,$\pm$\,0.10\\
\hline
070  &O9.7 II    &NGC~2060   &32.0\,$\pm$\,1.0 &4.00\,$\pm$\,0.10 &4.01 &5.64 &$-$13.35\,$\pm$\,0.15&126       &277 &7.00\,$\pm$\,0.20\\
077  &O9.5: IIIn &NGC~2060   &34.0\,$\pm$\,1.5 &4.20\,$\pm$\,0.15 &4.24 &5.20 &$<-$13.50        &225 &275 &7.10\,$\pm$\,0.30\\
080  &O9.7 II-III((n))&NGC~2060 &32.0\,$\pm$\,1.5 &3.85\,$\pm$\,0.15 &3.89 &7.55 &$<-$13.50     &190       &275 &7.00\,$\pm$\,0.20\\
091  &O9.5 IIIn  &NGC~2060  &32.0\,$\pm$\,1.5 &3.80\,$\pm$\,0.15 &3.88 &7.98 &$<-$13.50         &308       &275 &7.10\,$\pm$\,0.30\\
113   & O9.7 II/B0 IV? &NGC~2060 &32.0\,$\pm$\,1.0 &4.25\,$\pm$\,0.10 &4.25 &5.19 &$-$12.89\,$\pm$\,0.12 &\o35    &275 &7.00\,$\pm$\,0.15\\
128  &O9.5 III:((n))  &NGC~2060 &33.0\,$\pm$\,1.0 &4.15\,$\pm$\,0.15 &4.17 &5.01  &$<-$13.28     &180       &270 &7.00\,$\pm$\,0.20\\
131  &O9.7            &NGC~2060 &32.8\,$\pm$\,1.0 &4.38\,$\pm$\,0.10 &4.39 &4.49  &$-$12.87\,$\pm$\,0.3 &124    &275 &7.10\,$\pm$\,0.20\\
141  &O9.5 II-III((n))&NGC~2060 &32.3\,$\pm$\,1.0 &4.20\,$\pm$\,0.15 &4.21 &8.44  &$-$12.87\,$\pm$\,0.20&160    &275 &7.00\,$\pm$\,0.20\\
188* &O9.7: III       &NGC~2060 &34.0\,$\pm$\,2.0 &4.40\,$\pm$\,0.10 &4.40 &6.42  &$-$13.22\,$\pm$\,0.31&100&275 &7.00\,$\pm$\,0.20\\
192* &O9.7 II/B0 IV   &NGC~2060 &31.7\,$\pm$\,2.0 &4.13\,$\pm$\,0.10 &4.13 &4.85  &$-$13.05\,$\pm$\,0.17& \o40    &270 &7.10\,$\pm$\,0.20\\
205  &O9.7 II((n))/B0 IV((n))   &NGC~2060 &31.0\,$\pm$\,1.0 &4.30\,$\pm$\,0.15 &4.31 &6.29  &$<-$13.50       &158     &265 &6.90\,$\pm$\,0.20\\
207   &O9.7 II((n))    &NGC~2060 &31.4\,$\pm$\,1.0 &4.20\,$\pm$\,0.15 &4.21 &5.76 &$-$13.20\,$\pm$\,0.3 &160    &270 &7.10\,$\pm$\,0.20\\
210  &O9.7 II-III((n))&NGC~2060 &32.0\,$\pm$\,1.0 &4.00\,$\pm$\,0.15 &4.02 &6.47 &$-$12.78\,$\pm$\,0.2 &162    &275 &7.10\,$\pm$\,0.10\\
\hline
104 & O9.7 II-III((n))&Field &32.0\,$\pm$\,1.5 &4.10\,$\pm$\,0.15 &4.14 &5.10  &$-$13.50\,$\pm$\,0.30 &198    &260 &7.00\,$\pm$\,0.20\\
226  &O9.7 III        &Field  &33.0\,$\pm$\,1.0 &4.20\,$\pm$\,0.10 &4.20 &5.31 &$-$13.50\,$\pm$\,0.3  &\o64     &190 &7.10\,$\pm$\,0.10\\
328  &O9.5 III(n)     &Field  &33.6\,$\pm$\,1.5 &4.15\,$\pm$\,0.15 &4.19 &5.12 &$<-$13.28         &240    &300 &7.00\,$\pm$\,0.20\\
370  &O9.7 III        &Field  &32.0\,$\pm$\,1.0 &4.00\,$\pm$\,0.10 &4.00 &5.88 &$<-$13.47         & \o75&230&7.00\,$\pm$\,0.15\\
495  &O9.7 II-IIIn    &Field  &32.0\,$\pm$\,1.5 &4.20\,$\pm$\,0.15 &4.22 &6.44 &$-$13.50\,$\pm$\,0.2  &218   &270 &7.00\,$\pm$\,0.20\\
574  &O9.5 IIIn       &Field  &32.0\,$\pm$\,1.5 &4.10\,$\pm$\,0.15 &4.17 &5.17 &$<-$13.38         &270   &270 &6.90\,$\pm$\,0.20\\
753  &O9.7 II-III     &Field  &32.5\,$\pm$\,1.0 &4.00\,$\pm$\,0.10 &4.00 &7.74 &$<-$13.50  &{\it \o40}  &270 &7.20\,$\pm$\,0.10\\ 
787$^{+}$  &O9.7 III  &Field  &32.8\,$\pm$\,1.0 &4.35\,$\pm$\,0.10 &4.35 &5.77 &$-$12.76\,$\pm$\,0.2  &{\it \o40}  &275 &7.00\,$\pm$\,0.20\\
843  &O9.5 IIIn       &Field  &32.0\,$\pm$\,1.5 &3.90\,$\pm$\,0.15 &4.02 &6.03 &$<-$13.50         &318       &275 &7.10\,$\pm$\,0.30\\
\hline
\end{tabular}
\tablefoot{(1) \loggc\,$=$\,log[$g$\,+\,(vsin {\it i})$^{2}$/\Rstar], see e.g. \citet{repo}; (2) Adopted from RA17; 
(3) Italicised values differ from those from RA17 by more than 1$\sigma$; 
($\ast$) objects with asymmetric and blueshifted \heiic; 
($+$) Macroturblence of \vmac\,$=$\,45\,\kms was needed to reproduce the observed line profiles.}
\end{table*}
\end{center}
}

-- {\it Microturbulence (\vmic):} The classical approach to estimating
microturbulent velocities is to eliminate trends in the derived
abundance versus equivalent width of lines from a given ion.  For B-type
stars the \ion{Si}{iii} multiplet at $\sim$4560~\AA\ has been
repeatedly used for this purpose (see e.g.  \citealt{dufton05},
\citealt{hunter07}, \citealt{fraser10}).

Unfortunately, for the temperature regime considered here, the
Si~{{\scriptsize III} lines are weak and lie on or are close to the
  linear part of the curve of growth (i.e. their line strengths are
  relatively insensitive to changes in \vmic). The generally low S/N
  ratio of our spectra (recalling that most were rated BBB in
  Table~\ref{sample}), combined with the relatively rapid rotation
  rates of our targets, further complicates the situation as it makes
  it difficult to robustly constrain \vmic from the line-profile fits.
  Indeed, \citet{carolina14,carolina17} noted that the $\chi^{2}$
  distribution was degenerate in \vmic\ for many of their targets, and
  RA17 also found that the $\chi^{2}$ value for many of their stars
  was not significantly influenced by the specific values of \vmic, as
  long as it was below 30\,\kms (Ram\'irez-Agudelo, priv. comm.).

With this in mind, we used our model grid to put initial constraints on 
the microturbulent velocity, in parallel to those on \Teff, \logg, \Yhe, 
\epsSi\, and log$Q$. In particular, we 
found that models with \vmic\ = 5~\kms\ (in the formal integrals) 
most closely reproduced the profiles of strategic HHeSi lines 
in the spectra of our targets. To simplify the analysis and to 
reduce the computational effort, we did not attempt to improve 
the quality of the fit by iterating on this estimate, but adopted 
it as a final solution for all sample stars. This might influence the
estimated chemical abundances (see Sect.~\ref{Si}) but we 
consider this approach as justified given the difficulties of
actually deriving \vmic\ for stars at the cooler edge of the
O star temperature regime.

\smallskip

-- {\it Helium abundance (\Yhe):}  In the low-luminosity
high-gravity O star regime, He~{\scriptsize I} lines are
almost insensitive to changes in \Teff, \logg,\ and \Mdot, and can
therefore be used to estimate the helium abundance (provided \vmic\ is
independently fixed, see above). However, this approach was not
possible for some of our sample due to nebular contamination in some
of the most important He~{\scriptsize I} lines.

In contrast, the He~{\scriptsize II} lines ($\lambda\lambda$4200,
4541) do not suffer from nebular contamination in our targets and are
also insensitive to changes in \Mdot and \vmic\ (see Fig.~A.1. and
\citealt{VH00}). Although the intensity of He~{\scriptsize II} has a
strong dependence on both \Teff\ and \logg, these two parameters were
derived independently from the Si ionisation balance and the wings of
\Hdelta\ and \Hgama. We therefore used the He~{\scriptsize II} lines to
investigate \Yhe, with He~{\scriptsize I} as an additional check where
possible.

Good fits to He~{\scriptsize II}~$\lambda\lambda$4200, 4541,  in parallel to 
those to H and Si, were obtained for all stars using models with
\Yhe\,$=$\,0.1\,$\pm$\,0.02. When the He~{\scriptsize I} lines were
not influenced by nebular emission, they were also reproduced well by
the models. As expected, when the He~{\scriptsize I} profiles were
contaminated, the models over-predicted the absorption intensities.

\smallskip

-- {\it Silicon abundance (\epsSi):} This was constrained from the
best fit to selected Si lines. For late O-type stars, the
Si~{\scriptsize IV} lines are less sensitive to changes in \Teff\
compared to Si~{\scriptsize III} (see Sect.~\ref{spec_class}), so we
used these lines as the primary abundance diagnostic. We gave more
weight to the Si~{\scriptsize IV}~$\lambda\lambda$4116, 4212
transitions in the analysis because \SiIVa\ is blended with weak metal
lines and is also sensitive to \vmic\ (see Fig.~A.1). For stars with
\vsini$\le$100~\kms, an error of $\delta$\epsSi\,$=$\,0.1-0.15~dex was
estimated from the best fit to the corresponding lines; for those with
\vsini$>$100~\kms, a more conservative error of $\pm$0.2-0.3~dex was
adopted to account for the larger uncertainty caused by the fast
rotation. We remind the reader that these abundances (and also those
for He) were estimated given our specific assumptions on \vmic, that is,
\vmic\,=\,10~\kms in the NLTE calculations and \vmic\,=\,5~\kms for
the formal integrals. As such, the derived abundances are only valid
for these specific values, and we refer to Sect.~\ref{Si_formation}
for further discussions on this issue.

\smallskip

-- {\it Projected rotational velocity and macroturbulence:} The
rotational properties of our targets were determined by \citet{R13}
and RA17 by two different methods: use of Fourier transforms (FT) to
differentiate between the contributions from rotation and
macroturbulent broadening, and direct comparison of observed and
synthetic H and He profiles (where the effects of macroturbulence were
neglected).

The quality of our spectra and the typically weak metal lines prohibit
effective use of the FT approach. Therefore, we employed the second
method to fix the rotational properties of our sample, adopting the
\vsini\ estimates from RA17 as the initial input to compare between
the observed spectra and our model profiles for selected H, He, and Si
lines.  The initial values were only modified (by more than 1$\sigma$)
for four stars to improve the fits (VFTS\,412, 569, 753, and 787), which
we attribute to the different fitting approach here and the use of Si
lines as an additional indicator of \vsini\ compared to RA17. 

Comparing our \vsini\ estimates with those from the FT method by
\citet{R13} we do not see evidence of significant or systematic
offsets that might indicate additional broadening from macroturbulence. 
This is consistent with similar findings for Galactic stars from 
\citet{sergio17} which indicate that low-luminosity, high-gravity O 
stars have maximum macroturbulent velocities below the detection 
threshold of the FT method (see, e.g. their Figs. 4 and 5).

\smallskip

-- {\it Radial velocities:} Weighted means of the multi-epoch \Vr
estimates from \citet{sana13} were used as a starting value to fit the
selected HHeSi lines in the combined spectra of our targets. The
majority of these values were adjusted (by less than 3$\sigma$) to
optimise the fits. We attribute these differences to the different
fitting approach and the use of Si lines as an additional \Vr
indicator  \citep[cf. the original analysis by][]{sana13}.

\section{Results from model atmosphere analysis}\label{results}

The main stellar and wind parameters are listed in Table~\ref{para}, 
along with their corresponding errors derived as described above. The 
column entries are as follows: (1) VFTS identifier; (2) spectral
classification from \citet{walborn14}; (3-5) derived effective
temperature and surface equatorial gravity (effective and Newtonian);
(6) stellar radius from RA17; (7) wind strength Q-parameter from this
study; (8) \vsini; (9) \Vr, and (10) \epsSi. A helium abundance of
\Yhe\,$=$\,0.1\,$\pm$\,0.02 was found for all objects.
\begin{figure}
{\includegraphics[width=9.1cm,height=6.2cm]{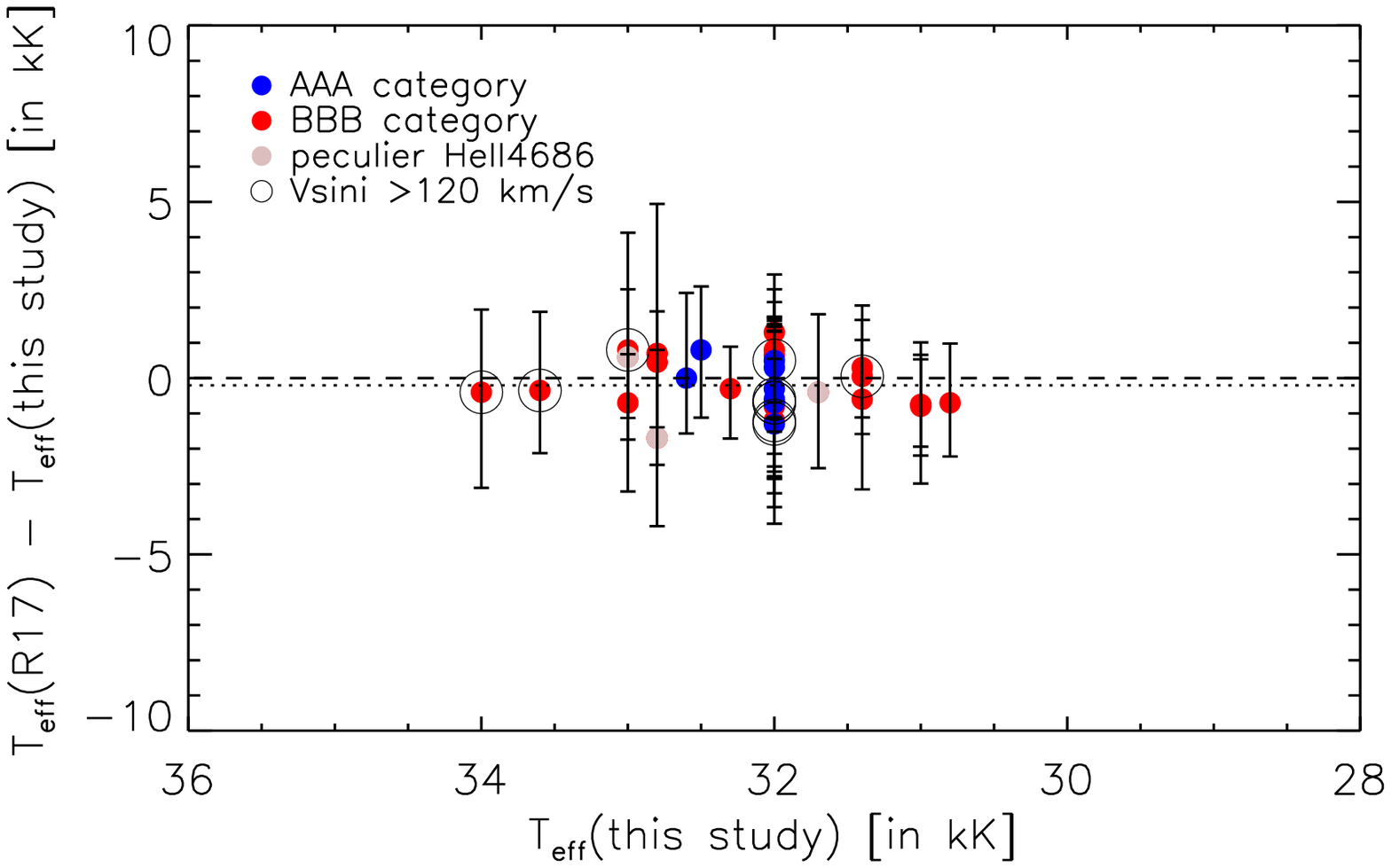}}\\
{\includegraphics[width=9.1cm,height=6.2cm]{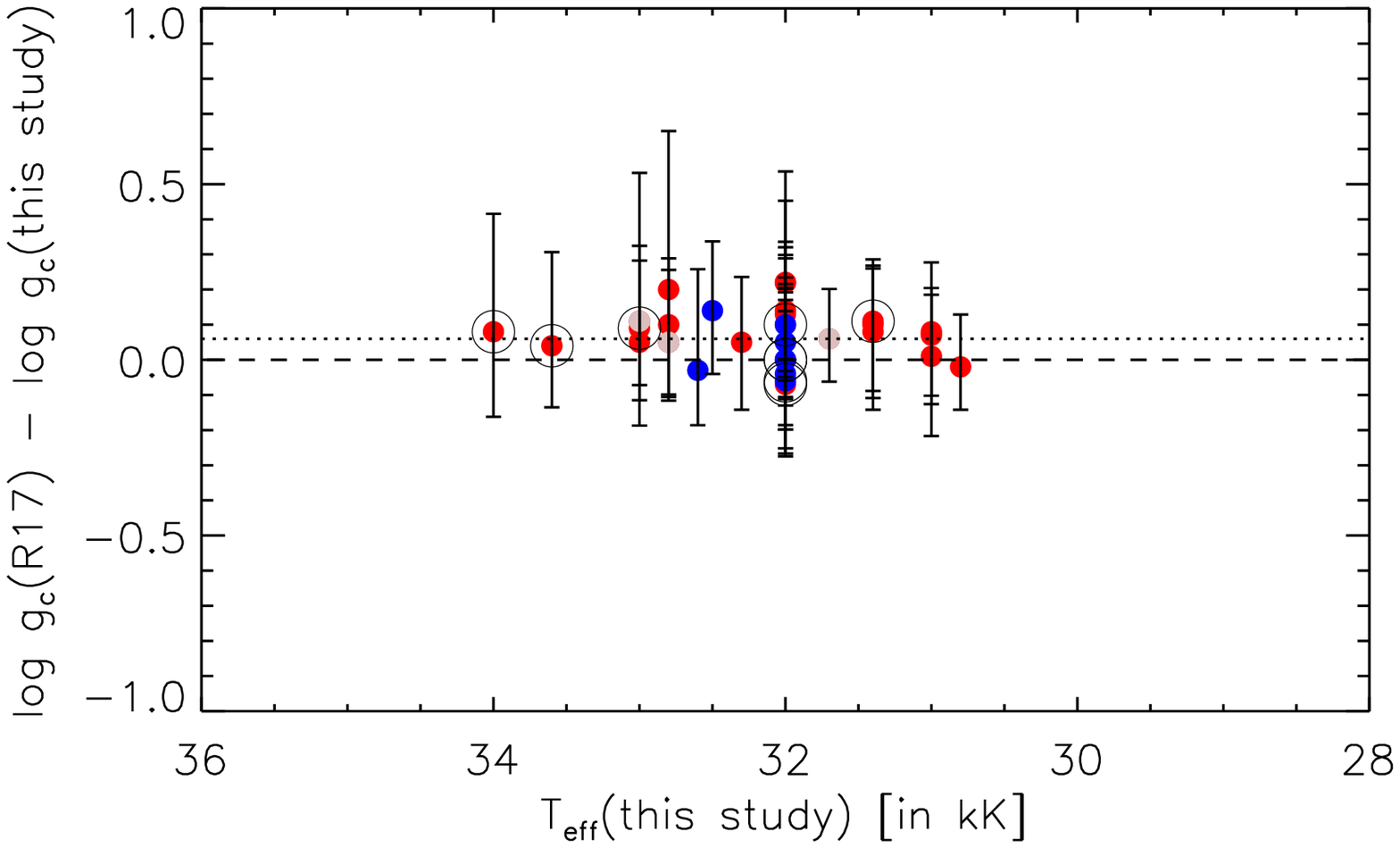}}\\
\caption{Differential effective temperatures (\Teff, upper panel) and
  equatorial surface gravities (corrected for centrifugal force,
  \loggc, lower panel), between those from RA17 and this study.
  Colours indicated in the legend distinguish between AAA- and
  BBB-rated spectra and those with peculiar \heiic\ (see text for
  details). Rapid rotators are highlighted with a large circle.
  Weighted mean and zero differences are shown by the dotted and
  dashed lines, respectively.}
\label{fig2}
\end{figure}

In the following sections we use these results compared to those from
RA17 to investigate four stellar parameters: \Teff, \loggc, $Q$, and
\epsSi.

\subsection{Effective temperatures and equatorial surface gravities}\label{TN-TR17}

The differences between the current estimates of \Teff\ and \loggc\
(HHeSi FW models, by-eye fits) and those from RA17 (HHe FW
models, automated GA fits) are shown in Fig.~\ref{fig2} (as a
function of our \Teff\ estimates). There appears to be good agreement
between the two sets of results (with differences within the
individual uncertainties), with no obvious dependence on \Teff, data
quality (AAA vs. BBB ratings), or rotational properties.

Weighted means of the differences (RA17\,$-$\,this study) are
$\Delta$\Teff\,$=$\,$-$0.14\,$\pm$\,0.31\,kK and
$\Delta$\loggc\,$=$\,0.06\,$\pm$\,0.04\,dex. These are smaller than
the typical errors in both studies ($\sim$1~kK in \Teff and 0.1~dex in
\logg), and are therefore statistically insignificant. The small differences
probably reflect the different methods used to determine stellar
temperatures (including the use of a single value for \vmic\ by RA17
in contrast to our two-value approach) and the weighting used here for
the wings of the Balmer lines.

\begin{figure}
{\includegraphics[width=9.1cm,height=6.2cm]{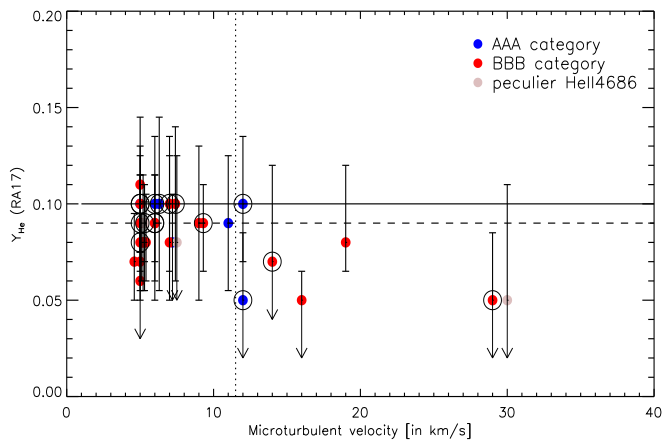}}\\
\caption{Results from RA17 for helium abundance (\Yhe) vs.
  microturbulent velocity for our sample stars (same symbols and colours
  as in Fig.~\ref{fig2}). Horizontal dashed and solid lines indicate,
  respectively, the expected value of \Yhe\ for unevolved massive
  stars in the LMC ($\sim$0.09, e.g. \citealt{Brott}) and the mean
  value of 0.1$\pm$0.02 derived from the current analysis (see
  Sect.~\ref{para_determinations} and \ref{he_abn} for details).}
\label{fig3}
\end{figure}

\subsection{Helium abundances}\label{he_abn}

As mentioned above, our analysis did not reveal evidence for a
significant underabundance of He in the sample stars. Indeed, a good
fit was obtained to the He lines for all targets (except for those
contaminated by relatively strong nebular emission; see
Sect.~\ref{para_determinations}) using models with \Yhe\,$=$\,0.1\,$\pm$\,0.02.
This differs from RA17 who obtained \Yhe\ ranging from 0.05 to 0.11.

In addition to the observational factors influencing the data (low
S/N, nebular contamination, or continuum dilution via an unresolved
companion or composite spectra), microturbulence might also
contribute to the large spread in \Yhe\ found by RA17.  Indeed, as
noted in Sect.~\ref{para_determinations}, \vmic\ was relatively
unconstrained in the RA17 analysis -- if overestimated the model
atmosphere fits might indicate lower \Yhe\ than the true value.

To investigate this further, the \Yhe--\vmic\ results from RA17 for
our targets are shown in Fig.~\ref{fig3} (in which the AAA- and
BBB-rated stars and those with peculiar \heiic\ are indicated),
revealing several important features:

\begin{itemize}
\item[i)] {Within the typical error of 0.03-0.04, the helium abundances
    derived by RA17 agree with our mean value and the \Yhe\,$=$\,0.09
    adopted by \citet{Brott} for non-evolved main sequence stars in
    the LMC.}

\item[ii)] {At \vmic\,$\ge$\,12\,\kms\ , RA17 found \Yhe\,$\le$\,0.08 for
    all six stars (where all but one are upper limits).  Such large
    values of \vmic\ would be unexpected for high-gravity, late O-type
    stars, suggesting that an overestimate of \vmic\ probably
    contributed to the low \Yhe\ estimates from the HHe model GA fits.}

\item[iii)] {At \vmic\,$<$\,12\,\kms\ RA17 found a range of \Yhe\ for
    similar values of \vmic. We interpret this spread as probably
    arising from the above observational factors (and noting that
    seven of the eight stars with \Yhe\,$<$\,0.08 were BBB-rated).}

\item[iv)] {Fast rotation (indicated by the additional large circles in
    Fig.~\ref{fig3}) does not appear to play a decisive role in the
    low helium abundances from RA17.}
    
\end{itemize}

\subsection{Wind strength parameter $Q$}\label{q_par}

Our 
estimates of log\,$Q$ spanned from $-$13.50 to $-$12.57\,dex,
comparable to the range of results from RA17
($-$13.84\,$<$\,log\,Q\,$<$\,$-$12.54). The differences between the
sets of results are smaller than the individual uncertainties, with no
clear evidence for a significant systematic offset
($\Delta$\,log\,$Q$\,$=$\,$-$0.06\,$\pm$\,0.37\,dex). The new results
therefore support the log\,$Q$ estimates from RA17, confirming that our
targets possess weak stellar winds, typical for low-luminosity O-type
dwarfs in 30 Dor
\citep[$-$13.5\,$\le$\,log\,Q\,$\le$\,$-$12.5][]{carolina17}.

\subsection{Silicon abundances}\label{Si}

Silicon abundances (and uncertainties) for our sample are shown in
Fig.~\ref{fig4} and listed in Col.~11 of Table~\ref{para}. The
estimates range from \epsSi\,$=$\,6.9 to 7.1, with only one star
(VFTS\,753) with \epsSi\,$=$\,7.2. This spread is comparable to the
typical error on our \epsSi\ determinations and might therefore be
important (see below). Averaging within each of the three spatial
subgroups, we found weighted mean abundances of 7.03\,$\pm$\,0.04 for
the stars in NGC~2070, 7.04\,$\pm$\,0.05 for those in NGC~2060, and
7.07\,$\pm$\,0.06 for those in the nearby field (where the quoted
uncertainties are the errors on the means). The differences between
these estimates are smaller than the corresponding errors and are thus
insignificant.
Comparing our estimates with that for the photospheric abundance of Si 
in the Sun (7.51$\pm$0.02 dex, see \citealt{asplund09}), we obtain 
a mean [Si/H] for our sample stars of $-$0.46$\pm$0.04.  

\begin{figure}
{\includegraphics[width=9.1cm,height=6.2cm]{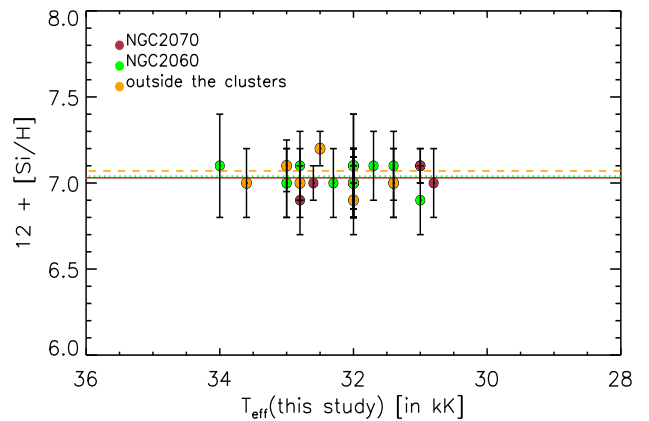}}\\
\caption{Si abundances and their uncertainties for our sample stars.
  Membership of the two main clusters in 30~Dor is colour-coded as
  indicated.  Weighted means for NGC~2070, NGC~2060, and field stars
  are shown by the horizontal solid, dotted, and red-dashed lines,
  respectively. We note that the parameters for some stars are the same, 
hence the number of displayed points appears smaller than the total number
in the sample.}
\label{fig4}
\end{figure}

To our knowledge, these are the first Si abundances for O-type stars
in the LMC and therefore we compare them with results for early B-type 
stars in LMC clusters from \citet{korn00,korn02,korn05}, \citet[][hereafter
T07]{trundle07}, and \citet[][hereafter H07]{hunter07}. The mean Si
abundances from these studies are given in Col.~4 of Table~\ref{Siabn}. 
Compared to the H07 and T07 results, our (weighted) mean abundances 
are 0.2\,dex lower, but agree well with those from \citeauthor{korn00}

The series of papers by \citeauthor{korn00} used the NLTE
line formation codes {\sc detail} and {\sc surface} \citep{BG85} in
combination with plane-parallel, fully line-blanketed LTE Kurucz model
atmospheres ({\sc atlas}{\scriptsize 9,} \citealt{kurucz93}) and
high-resolution spectroscopy to determine surface chemical abundances
for limited samples of main sequence B stars in the young clusters
NGC~1818 and NGC~2004. 

The H07 and T07 studies employed optical spectroscopy from the
FLAMES~I survey \citep{evans05,evans06} and the Queen's University
Belfast (QUB) grid of NLTE, plane-parallel, static (i.e. no wind)
models calculated with {\sc tlusty} and {\sc synspec}
\citep{ryans03,dufton05}. These latter studies investigated the 
surface chemical abundances of main sequence B-stars in the young 
star-forming region N11 (H07) and also in NGC~2004 (T07).  We add 
that the same QUB {\sc tlusty} grid (and similar methods to H07 and 
T07) was also used by \citet{mcevoy15} and 
\citet[][hereafter D18]{dufton18} to estimate the properties of a
large sample of B-type stars from the VFTS, although we note that in
these studies the Si abundances were fixed at the value derived by H07
and the Si~{\scriptsize III} triplet ($\lambda\lambda$4553, 4568,
4575) was then used to estimate the microturbulence.

\begin{table*}
\begin{center}
\caption[]{Mean silicon abundances for OB-type stars in the LMC from
    the current study and the literature. The quoted uncertainties
    correspond to either the error of the mean (present study) or the
    standard deviation of the individual results.}\label{Siabn}
\begin{tabular}{lllccl}
\hline\hline
\multicolumn{1}{l}{Cluster/Region}
&\multicolumn{1}{l}{Spectral type}
&\multicolumn{1}{l}{Models }
&\multicolumn{1}{c}{Mean \epsSi$^{(a)}$}
&\multicolumn{1}{c}{\logg~(cm\,s$^{-2}$)}
&\multicolumn{1}{l}{Comments}\\
\hline
NGC 2070     & Late O & {\sc fastwind} HHeSi & 7.03\,$\pm$\,0.04 (10) & 3.8-4.4\o & This study \\
NGC 2060     & Late O & {\sc fastwind} HHeSi & 7.04\,$\pm$\,0.05 (13) & 3.8-4.4\o & This study \\
30~Dor field & Late O & {\sc fastwind} HHeSi & 7.07\,$\pm$\,0.06 (9)\o  & 3.8-4.4\o & This study \\
\hline
N11      & Early to mid B & {\sc tlusty}/{\sc synspec} & 7.19\,$\pm$\,0.07 (30) & 2.1-4.2\o  & \citet{hunter07} \\
NGC 2004 & Early to mid B & {\sc tlusty}/{\sc synspec} & 7.21\,$\pm$\,0.03 (23) & 1.9-4.15 & \citet{trundle07} \\
NGC 2004 & Early B & {\sc atlas}{\scriptsize 9}/{\sc detail}/{\sc surface} & $\phantom{^{(b)}}$7.00\,$\pm$\,0.16 (6)$^{(b)}$\o & 3.1-4.35 & \citet{korn00,korn02} \\
NGC 1818 & Early B & {\sc atlas}{\scriptsize 9}/{\sc detail}/{\sc surface} & $\phantom{^{(b)}}$7.03\,$\pm$\,0.12 (2)$^{(b)}$\o &2.8-4.0\o  & \citet{korn00,korn02} \\
NGC 2004 & Early B$^{(c)}$ & {\sc atlas}{\scriptsize 9}/{\sc detail}/{\sc surface}  &7.07\,$\pm$\,0.06 (3)\o & 3.7-4.05 & \citet{korn05} \\
\hline               
\end{tabular}
\tablefoot{(a) Number of objects in parentheses; (b) Means
calculated from the individual estimates in the relevant studies;
(c) Relatively rapid rotators (\vsini\,$\sim$\,130\,\kms).}
\end{center}
\end{table*} 

There are several reasons that might contribute to explain the
systematic offset between the silicon abundance estimates obtained
here and those from H07 and T07, the most important being different
methodologies and/or atomic data (particularly for Si), the effects of
microturbulence, and real differences between the studied objects.

\subsubsection{Comparison of FASTWIND and TLUSTY results}\label{fw_tlusty}

\citet{mcevoy15} compared the physical parameters estimated for ten
early B-type supergiants and two O9.7 supergiants using the QUB {\sc
 tlusty} grid and HHe FW models. The results for \Teff, \logg, and
nitrogen abundances were generally in good agreement between the two
methods (see their Table~6), as also found for two B supergiants in
the Small Magellanic Cloud by \citet{dufton05}.

As a first comparison of results from the different codes for the
stars with unexpected gravities, RA17 compared their results from FW
with preliminary estimates from the QUB {\sc tlusty} grid for six of
the O9.7-type giants/bright giants in the current sample (VFTS\,113,
192, 226, 607, 753, and 787). The FW results showed that these latter 
objects were typically slightly hotter, with a mean temperature 
difference of nearly 2\,000\,K.

Briefly, the two codes and analysis methods broadly lead to
consistent results, although small systematic differences in \Teff\
and \logg\ cannot be excluded. To investigate this further we 
analysed three of the early B-type dwarfs (VFTS\,119, 313, 623) from
D18 with the new HHeSi FW models.  Provided \vmic\ for the formal 
integrals agrees to within $\pm$1~\kms, the two methods give 
estimates of \Teff\ and \logg\ that agree within the respective
uncertainties, although FW again leads to slightly hotter temperatures
($\Delta$\Teff[FW\,$-$\,{\sc tlusty}]\,$=$\,0.50\,$\pm$\,0.87~kK) and
marginally larger gravities ($\Delta$\logg[FW\,$-$\,
{\sc tlusty}]\,$=$\,0.03\,$\pm$\,0.03\,dex).  Interestingly, the FW
analysis also resulted in lower estimates of \epsSi\ than those from
\citeauthor{dufton18}, where $\Delta$\epsSi[FW\,$-$\,{\sc
  tlusty}]\,$=$\,$-$0.11\,$\pm$\,0.03.

While differences in the atomic data/models for Si cannot be excluded 
as the cause, we suggest that differences in the microturbulent
velocity may contribute to the small 
offset between the FW and {\sc tlusty} solutions for \Teff, \logg\ and 
\epsSi\ for late O- and early B-type stars in the LMC.

Indeed, as outlined in Sect.~\ref{para_determinations}, for
computational reasons we adopted microturbulent velocities of 10 and
5\,\kms\ in the NLTE and formal integral calculations, respectively,
for the analysis of our sample. In contrast, the studies of B-type
stars by H07, T07, \citet{mcevoy15}, and D18 adopted one single value 
for this quantity, where it was estimated from a comparison with 
observations via two methods: the classical approach based on curve-of-growth 
analysis of the Si~{\scriptsize III} triplet ($\lambda\lambda$ 4553, 4568, 4675), 
and an alternative approach where \vmic\ was adjusted for each target so 
that the derived \epsSi\ was consistent with the corresponding derived 
(H07, T07) or adopted \citep{mcevoy15,dufton18} cluster average.
\begin{figure}
{\includegraphics[width=9.1cm,height=6.2cm]{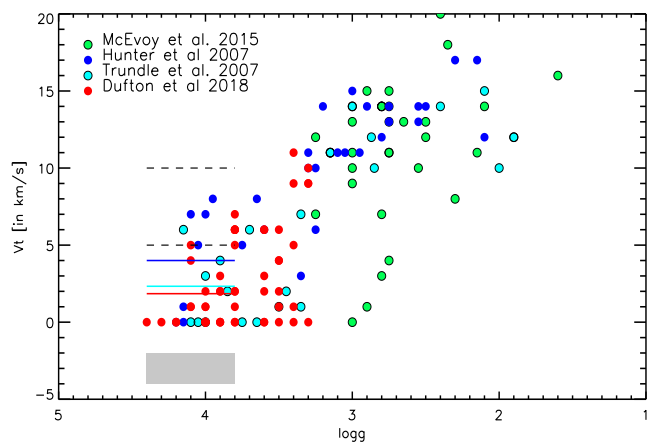}}
\caption{Estimated microturbulent velocities (\vmic) vs. equatorial
  surface gravities (\logg) from published studies of B-type stars in
  the LMC (colour-coded as in the legend). The grey rectangle
  highlights the range of gravities for our targets and the horizontal
  lines indicate the mean microturbulence for stars in that range from
  the published studies.  See text for further details.}
\label{fig5}
\end{figure}

The \vmic\, estimates from the second approach are shown in Fig.~\ref{fig5} 
as a function of \logg. Typical uncertainties are 
$\Delta$\logg\,$=$\,$\pm$0.1\,dex and
$\Delta$\vmic\,$=$\,$\pm$3-5\,\kms.  As already noted in previous
studies (e.g. \citealt{daflon04}, H07, \citealt{mokiem07}), for
low-gravity stars (\logg\,$<$\,3.2\,dex) there is a clear trend of
increasing \vmic\ with decreasing gravity.  For the high-gravity stars
(\logg\,$>$\,3.8\,dex) there is little evidence of a correlation.

To compare the microturbulent velocities adopted in the present 
study with the estimates in Fig.~\ref{fig5}, we calculated mean values 
from the published samples, including only those stars with gravities 
that span the same range as our targets (indicated by the grey rectangle). 
These means are indicated in the figure using horizontal solid lines 
and the same colours as in the legend: H07 (dark blue), $\overline{v_{\rm
    mic}}$\,$=$\,4.0\,$\pm$\,3.6\,\kms (7 stars); T07 (cyan),
$\overline{v_{\rm mic}}$\,$=$\,2.3\,$\pm$\,2.1\,\kms (6 stars);
\citet[][red]{dufton18}, $\overline{v_{\rm
    mic}}$\,$=$\,1.9\,$\pm$\,2.3\,\kms (34 stars).

Our adopted values of 10 and 5\,\kms\ (indicated by the dashed
horizontal lines in Fig.~\ref{fig5}) are, in part, significantly
larger than the corresponding results from H07, T07, and D18. 
With all other parameters fixed, higher values of \vmic\ would be 
expected to yield lower Si abundances, suggesting that differences 
in the microturbulent velocities adopted for the NLTE and the formal 
integral calculations might contribute to the small offset between 
the FW and {\sc tlusty} solutions for \Teff, \logg,\ and \epsSi\ for 
late O and early B-stars in the LMC (see following section).

\subsubsection{Effect of microturbulence on Si line diagnostics:
Theoretical considerations}\label{Si_formation}

The effect of microturbulence on the line formation calculations of H
and He lines in the optical spectra of O and early-B stars of solar
metallicity was explored in detail by \citet{SH98} using the {\sc
  tlusty/surface} codes, and by \citet{VH00} and \citet{LRL02}
using the {\sc detail/surface} codes. The main outcome of these
analyses was that only the He~{\scriptsize I} and \heiic\ lines are
affected by changes in this parameter (both in the NLTE and
the formal integral), and that the impact of these changes on the
derivation of \Teff, \logg, and \Yhe\, is small or comparable to the
adopted uncertainties.

To investigate the effect of microturbulence on the Si lines used in
the present study (see Table~\ref{diagnostic_lines}), we used
models with typical parameters for our targets: \Teff\,$=$\,32\,kK,
\logg\,$=$\,4.1~dex, \Yhe\,$=$\,0.1,  \epsSi=7.0\,dex, and log\,$Q$\,$=$\,$-$13.5. In
order to disentangle the relative influence of changes in the NLTE
occupation numbers and the formal integral, two sets of computations
were carried out: in the first set, the value of \vmic\ used in the
NLTE calculations, that is, \vmic(NLTE), was varied while that for the formal
integral was fixed at 5~\kms (see Sect.~\ref{para_determinations}); in
the second set, \vmic\, for the formal integral was varied whilst that
for the NLTE computations was fixed at 10~\kms, as adopted by us
throughout our analysis of the current sample (see Sect.~\ref{grid}).

Results from the first set of computations are illustrated in
Fig.~\ref{A1} which compares selected HHeSi lines calculated with
\vmic(NLTE)\,=\,5, 10, and 15~\kms. By visual inspection of the
profiles, only the Si~{\scriptsize III} lines are significantly
sensitive to changes in this quantity (11 to 17\% increase in EWs when
increasing \vmic(NLTE) from 5 to 15~\kms). In contrast, the Balmer
lines are not affected, and the He and Si~{\scriptsize IV} lines are
only marginally affected (with typical increases in EW of $\le 4\%$
for He and between 6 and 7\% for Si~{\scriptsize IV}).

The impact of these changes on the outcome of our analysis is minimal,
and in particular smaller than the typical uncertainties. Specifically, a
decrease from 10~\kms  (present study) to 5~\kms ( a value similar 
to $\overline{v_{\rm mic}}$  used by H07 
to analyse B-type dwarfs, see Fig.~5) leads to effective temperatures 
that are 300\,K cooler, logarithmic gravities that are 0.03\,dex 
lower, and Si abundances that are $\sim$0.05\,dex larger. These predictions are
qualitatively consistent with the differences between the FW and {\sc
  tlusty} analyses of common stars (see previous section), supporting
our suggestion that the \vmic\ value adopted in the NLTE calculation
can affect the outcome of model-atmosphere analyses when using the Si
lines as diagnostics of \Teff, \vmic, and \epsSi.  Similar calculations
using {\sc cmfgen} \citep{hm98} have shown that the formation of
specific infrared lines in O-type stars with very thin winds, namely
\Bralfa\ and \Pfgama, is also sensitive to the value of \vmic\ adopted
for the model-atmosphere calculations \citep{paco11}.

The effect of microturbulence on the emergent HHeSi profiles is
visualised in Fig.~\ref{A2}, where line profiles computed with
different \vmic\ values in the formal integral (5, 10, and 15~\kms)
are compared.  From these synthetic spectra, one can see that Si lines
are most sensitive to changes in this quantity, with Si~{\scriptsize
  IV} being more affected than Si~{\scriptsize III} (with typical EW
variations of 18-68\% for the former, and 7-34\% for the latter). For
the He lines, the situation is more complex, with the He~{\scriptsize
  I} lines showing modest variations, with a typical EW increase
between a few and 26\% (when varying \vmic(formal) between 5 and
15~\kms). The He~{\scriptsize II} lines at $\lambda\lambda$4200, 4542
are almost insensitive, while the Pickering line $\lambda$4686
exhibits a 13\% increase in EW.  An interesting point to be noted here
is that, as in the case established for early B-type stars (see
\citealt{LRL02}), He~I $\lambda$6678 \AA\, is most strongly affected
by changes in \vmic(formal). Therefore, also in the low-luminosity
high-gravity O star regime, this line cannot serve as a reliable
diagnostic to determine stellar properties.

Further test calculations showed that a decrease in \vmic(formal) from
5~\kms (present study) to 3~\kms (the average value used to analyse
B-type dwarfs by H07, T07, and \citealt{dufton18}, see Fig~\ref{fig5})
leads to an increase in \epsSi\ by $\sim$0.10\,dex (\Teff\ and \logg\
are not affected). Using the same version of the FW code,
\citet{gonzalez12} estimated that for O stars in the LMC, a 5\,\kms
decrease in \vmic\, for the formal integral leads to a 0.05-0.07\,dex
increase in nitrogen abundances.

In summary, for the Balmer and He lines the FW predictions outlined
above are in good qualitative agreement with similar findings from
\citet{SH98}, \citet{VH00}, and \citet{LRL02}; see also \citet{martins02}. For
Si, our studies and predictions have, for the first time, revealed
that in the low-luminosity, high-gravity O star regime, Si~{\scriptsize
  III} is more sensitive to changes in \vmic(NLTE) than
Si~{\scriptsize IV}, whereas the opposite effect (i.e.
Si~{\scriptsize IV} more strongly affected than Si~{\scriptsize III})
is found when \vmic(formal) is changed.  With regard to the
microturbulent velocities used in our analysis (see
Sect.~\ref{para_determinations}), variations of $\pm$5~\kms for the
NLTE and of $\pm$2~\kms for the formal integral calculations have no
significant consequences for the derivation of \Teff\ and \logg. In
contrast, \epsSi\ is expected to be significantly modified, with
variations of $\sim$0.05~dex for $\Delta$\vmic(NLTE)\,$=$\,5~\kms and by
$\sim$0.10~dex for $\Delta$\vmic(formal)\,$=$\,2~\kms.

The main implication of these theoretical considerations is that the
reconciliation of the \epsSi\ estimates derived for low-luminosity 
late O-type dwarfs (present study) with those derived for early-B stars 
(H07, T07) would require the atmospheres of the former to be (almost) 
free of microturbulent motions, at least in the line-forming region of 
Si (\vmic$<$5~\kms for both the NLTE and the formal integral calculations). 
However, such low values are not typical for O-stars in the Galaxy (e.g.
\citealt{martins12, martins15}, \citealt{mahy15}, \citealt{markova18})
or the Magellanic Clouds (see e.g. \citealt{bouret13}, \citealt{massey09, 
massey13}, \citealt{shenar15, shenar16}, \citealt{R18}).

Therefore, we conclude that while different assumptions in corresponding FW
and {\sc tlusty} analyses (particularly the microturbulent velocities
used in the NLTE and the formal integral calculations) can contribute,
it is unlikely that they alone can explain the differences in the
estimated Si abundances from H07 and T07 compared to our results.

\subsubsection{Revisting the low Si abundances for some LMC stars.}\label{low_Si}

Large spreads in individual Si abundances ($\Delta$\epsSi\ of up to
0.5-0.6\,dex) were reported by H07, T07, and D18 when
following the classical approach to determine \vmic\ for the formal
integral. These spreads were significantly reduced using the
alternative approach (adjusting this value to obtain \epsSi\ values
that are consistent with the corresponding cluster average; see
Sect.~\ref{fw_tlusty}) to determine \vmic. But even this approach
failed for a non-negligible number of objects (generally dwarfs in
NGC~2060 and NGC~2070) since the maximum possible Si abundance (for
\vmic\,$=$\,0\,\kms) turned out to be smaller than the corresponding
cluster averages of $\sim$7.2\,dex, by 0.2-0.3\,dex.

Following a similar approach, \citet{mcevoy15} found \vmic\,$<$\,10\,\kms\ 
for a number of their targets (see Fig.~\ref{fig5}). This is unexpectedly 
low for B-type supergiants, and in some cases deviates by more than 
3$\sigma$ from the apparent \logg--\vmic\ trend. If these values were 
increased to the values expected for their gravities, these would also lead 
to further B stars with \epsSi\,$<$\,7.2\,dex located in NGC~2060 
and NGC~2070.

If not due to limitations in the method and/or models, the above
discussion suggests that B stars with \epsSi\ as low as 6.9--7.0~dex
are present in the LMC, especially in the NGC~2060 and NGC~2070
clusters in 30~Dor. As for the late O-dwarfs in those clusters, a mean
\epsSi\,$=$\,7.05\,$\pm$\,0.03 was also derived in
the present study, leading us to speculate that the clusters in 30~Dor
might be somewhat more metal (or at least Si) deficient compared to
N~11 and NGC~2004.

The existence of a large spread in metallicity in the younger stellar
populations of the LMC is well known (see e.g.  \citealt{piatti19} and
references therein), but a careful reanalysis of the larger samples of
late-O and early-B stars observed in the VFTS and FLAMES~I survey is
required to more critically investigate the possibility of such
differences in metal enrichment.

\section{Spectral classification for late O-type stars}\label{spec_class}

From the above results it seems clear that our targets have physical
properties (particularly \Teff, \logg, \Yhe\ and log\,$Q$) similar to
those typical for O-type dwarfs in the LMC. With this in mind, we further
investigated the second hypothesis proposed in Sect.~\ref{intro}, namely 
that intricacies in spectral classification might be
responsible for the unexpected results from atmospheric analysis of
our targets with respect to their morphological types.

\subsection{Classification of the VFTS sample}

A new atlas for spectral classification of Galactic O-type spectra was
presented by \citet{sota11}, including updated spectral standards.
Among other refinements, the late-O/early-B spectral types were
redefined, including the introduction of the O9.7 spectral type at
lower luminosity classes (V through to III); previously, it had been
defined only for classes I and II (see Table~\ref{SpT}). The adopted
principal luminosity criterion for late O-types was the ratio of
\heiic/\heid\ (see Table~\ref{LC}), which is expected to decrease with
increasing luminosity due to infilling of \heiic\ by emission. This
contrasts with earlier studies that used the \SiIVa/\heia\ ratio as
the primary diagnostic \citep[e.g.][]{WF90}.

\begin{table}
\begin{center}
  \caption[]{Spectral type criteria at O8.5--B0 based on specific 
    line-strength ratios \citep{sota11}.}
\label{SpT}
\tabcolsep1.7mm
\begin{tabular}{lcc}
\hline\hline
\multicolumn{1}{l}{Spectral} &\multicolumn{1}{c}{\o He~{\scriptsize II}/He~{\scriptsize I}$^{(a)}$} &
\multicolumn{1}{c}{\o Si~{\scriptsize III}/He~{\scriptsize II}$^{(b)}$}\\
\multicolumn{1}{l}{Type} &\multicolumn{1}{c}{$\lambda$4542/$\lambda$4388} &\multicolumn{1}{c}{$\lambda$4553/$\lambda$4542}\\
\multicolumn{1}{c}{} &\multicolumn{1}{c}{$\lambda$4200/$\lambda$4144} &\multicolumn{1}{l}{}\\
\hline
O8.5       &$\ge$  &N/A\\
O9         &$=$    &$<<$\\
O9.5       &$\le$  &$<$\\
O9.7       &$<$    &$\le$ to $\ge$\\
B0         &$<<$   &$>>$\\
\hline              
\end{tabular}
\end{center}
\small{\bf Notes.} (a) primary criterion; (b) secondary criterion.\normalsize
\end{table}

\begin{table}
\begin{center}
\caption[]{Luminosity criteria for types O9--O9.7 \citep{sota11}.}\label{LC}
\tabcolsep2.5mm
\begin{tabular}{lcc}
\hline\hline
\multicolumn{1}{l}{Lum.} &\multicolumn{1}{c}{\o He~{\scriptsize II}/He~{\scriptsize I}$^{(a)}$} 
&\multicolumn{1}{c}{\o Si~{\scriptsize IV}/He~{\scriptsize I}$^{(b)}$}\\
\multicolumn{1}{l}{Class} &\multicolumn{1}{c}{$\lambda$4686/$\lambda$4713} &\multicolumn{1}{l}{$\lambda$4089/$\lambda$4026}\\
\hline
Ia       &$\sim$0      &N/A\\
Iab      &$\ll$ to $<$  &$\ge$ to $\le$\\
Ib       &$\le$        &$\le$\\
II       &$=$          &$<$\\
III      &$>$          &$<$ to $\ll$\\
V        &$\gg$         &$\ll$\\
\hline              
\end{tabular}
\end{center}
\tablefoot{(a) primary criterion; (b) secondary criterion.}
\end{table}

Extending these criteria to lower-metallicity environments such as the
LMC is not trivial. In addition to the Si lines weakening, the
strengths of the He lines are also expected to depend on the metal
content \citep[see][]{markova09}. To account for this
\citet{walborn14} classified each VFTS spectrum as if it were a
Galactic star using the criteria in Tables~\ref{SpT} and \ref{LC},
while also paying attention to differences between the estimated
absolute visual magnitudes and those expected given the
classification\footnote{Larger than expected luminosities could also
  arise from binarity as well as anomalous classifications. Although
  our sample is made up of single stars to the best of our current knowledge, an
  inability to obtain satisfactory fits for the selected H, He, and Si
  lines with a single set of parameters might indicate previously
  undetected binaries and/or composite spectra.}.

\begin{figure*}
{\includegraphics[width=6.1cm,height=4.5cm]{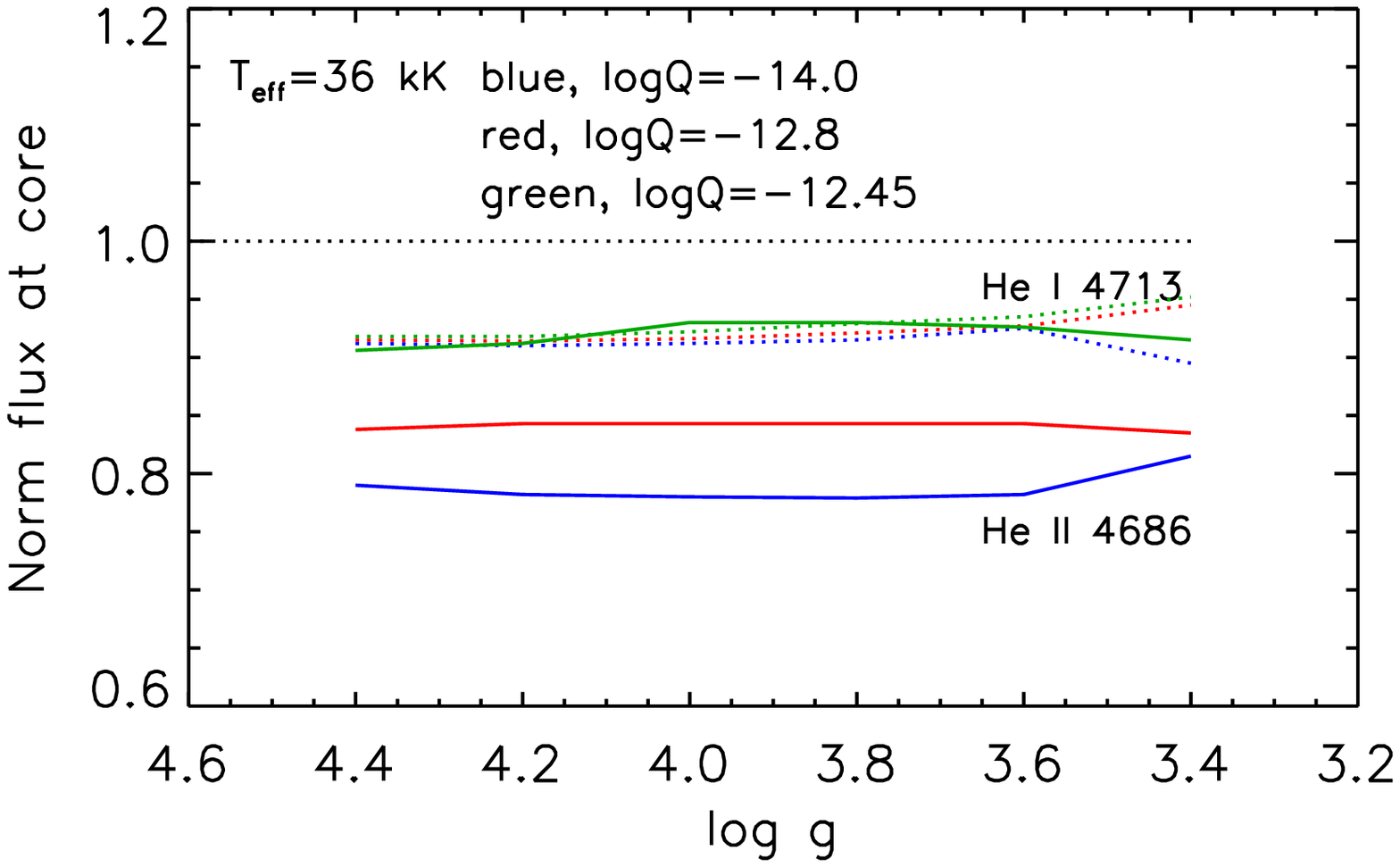}}
{\includegraphics[width=6.1cm,height=4.5cm]{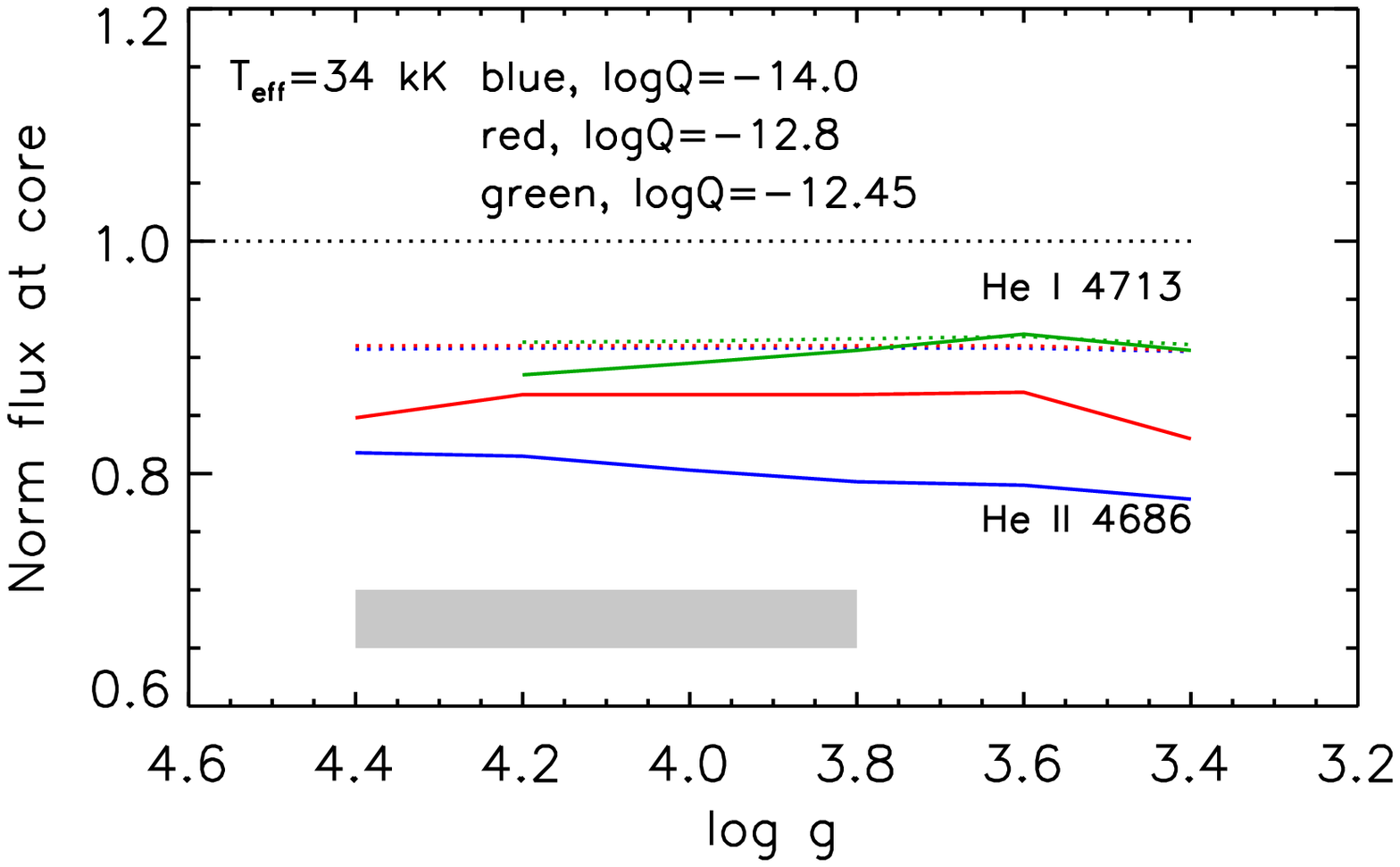}}
{\includegraphics[width=6.1cm,height=4.5cm]{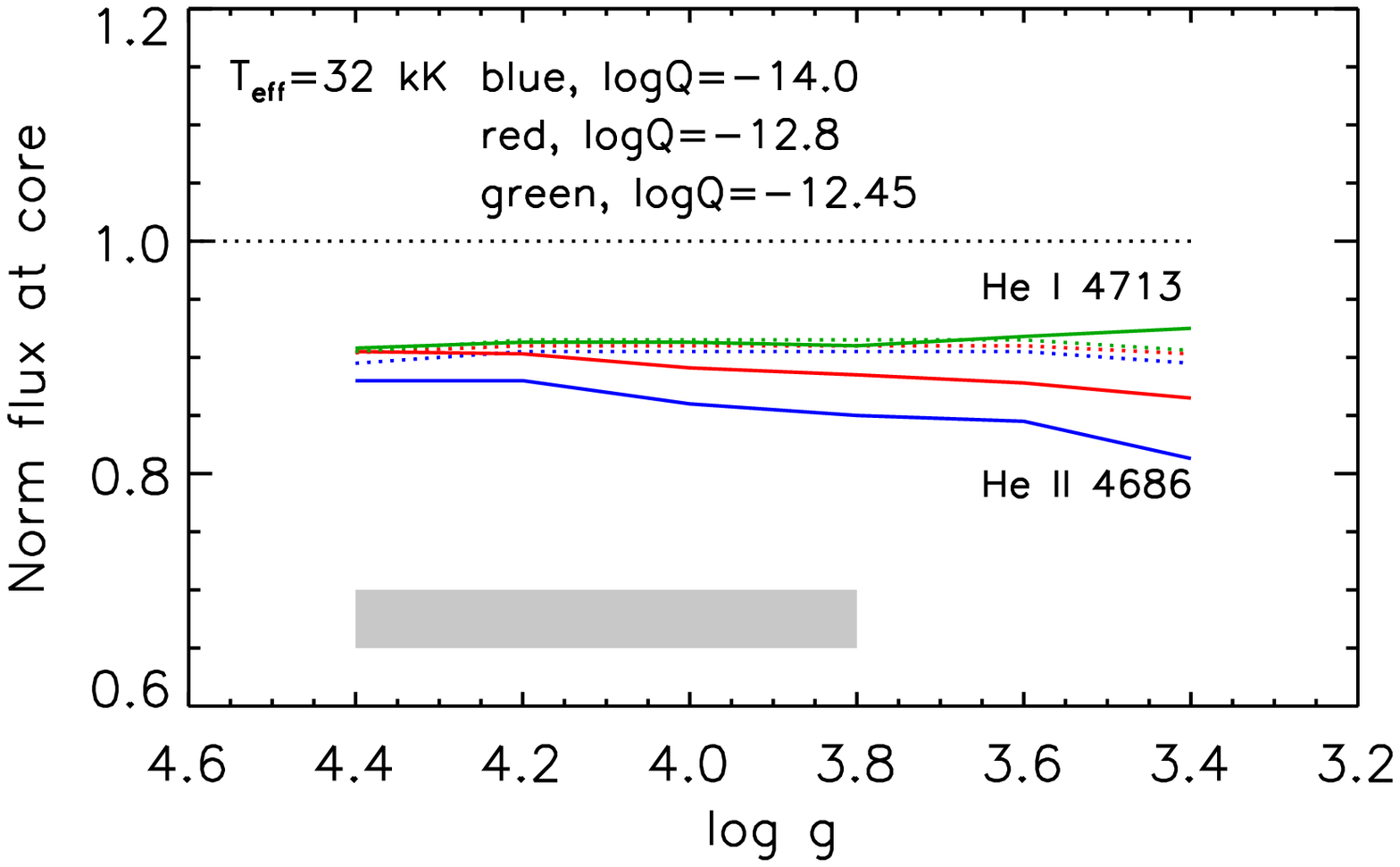}}\\
{\includegraphics[width=6.1cm,height=4.5cm]{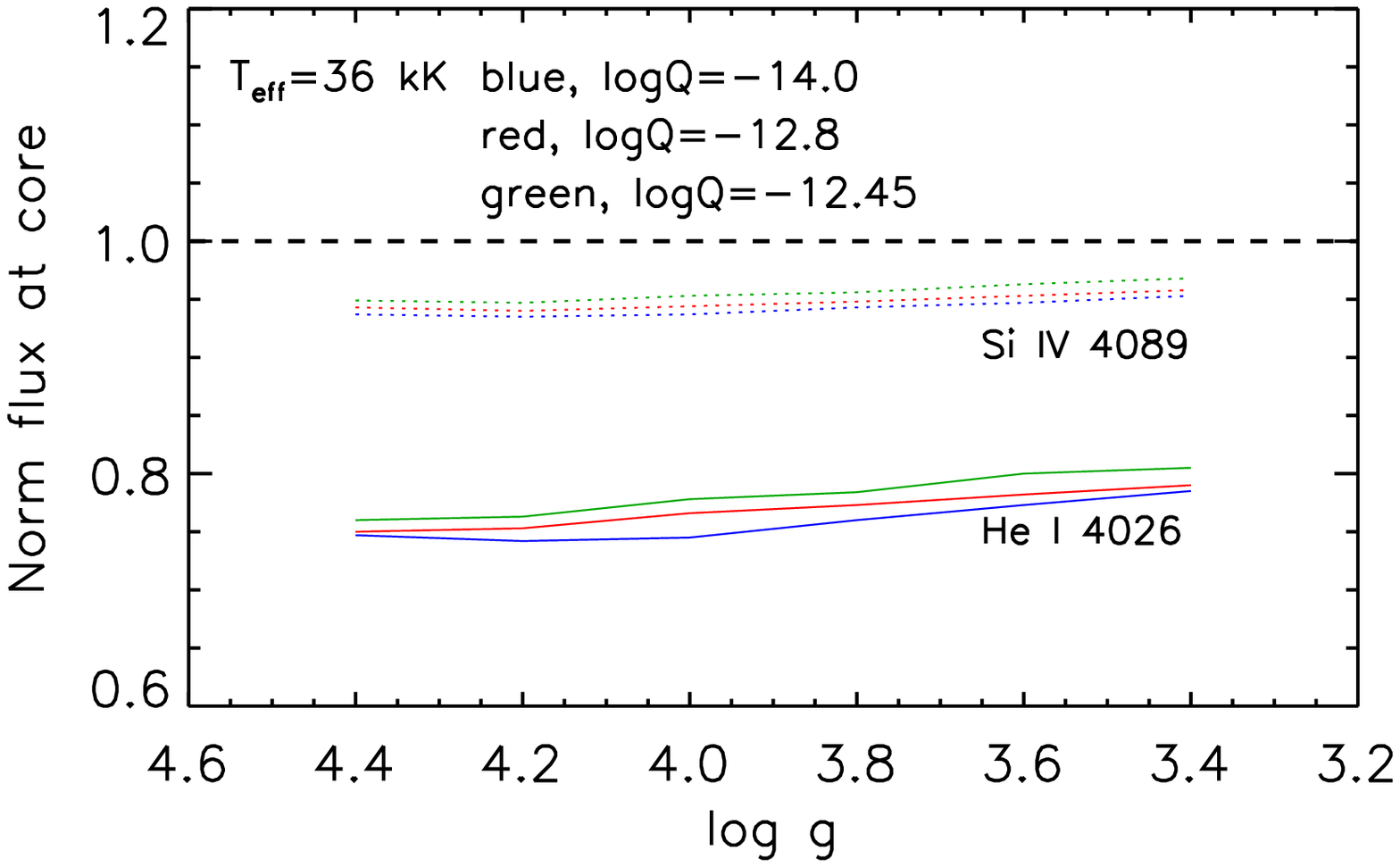}}
{\includegraphics[width=6.1cm,height=4.5cm]{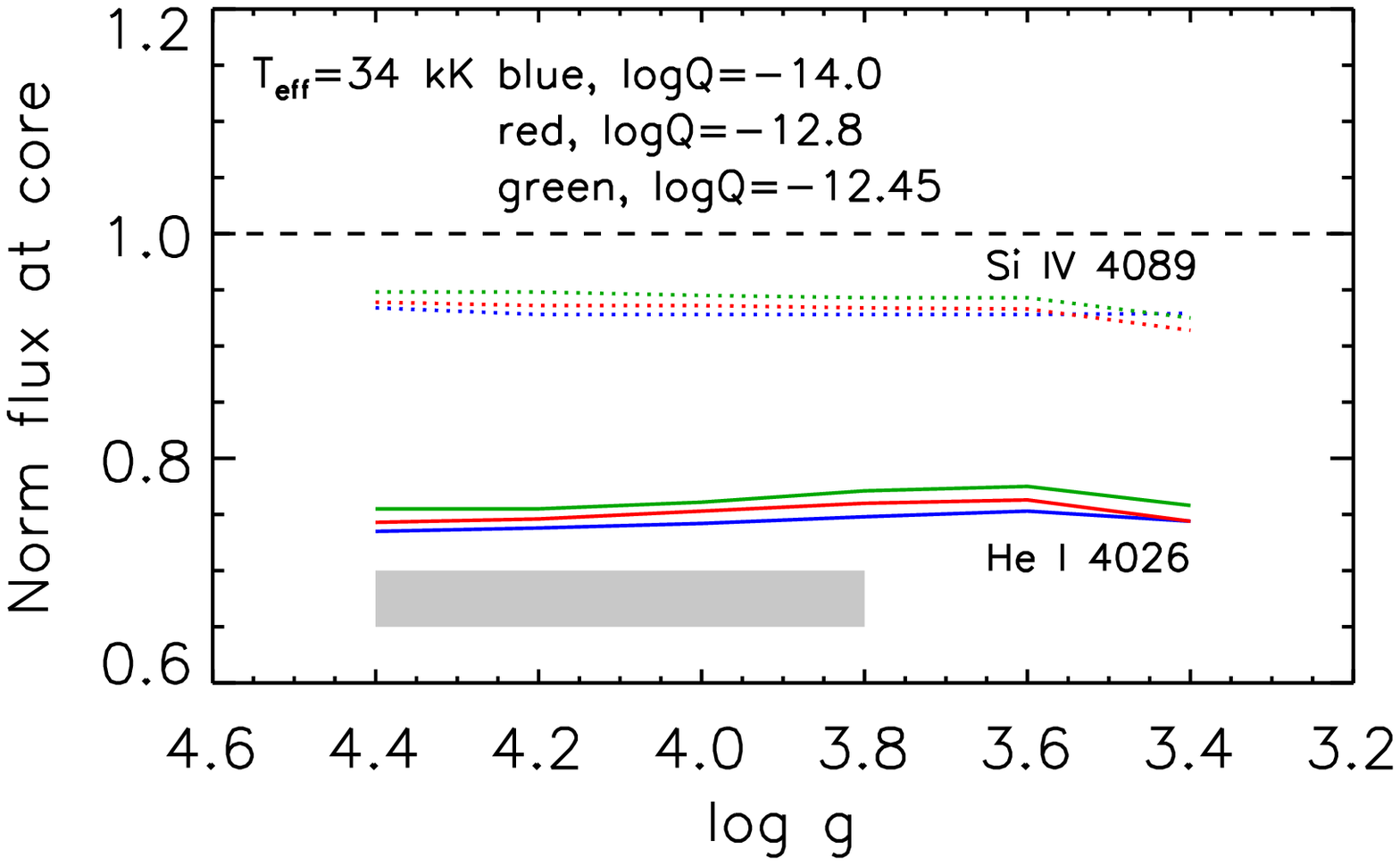}}
{\includegraphics[width=6.1cm,height=4.5cm]{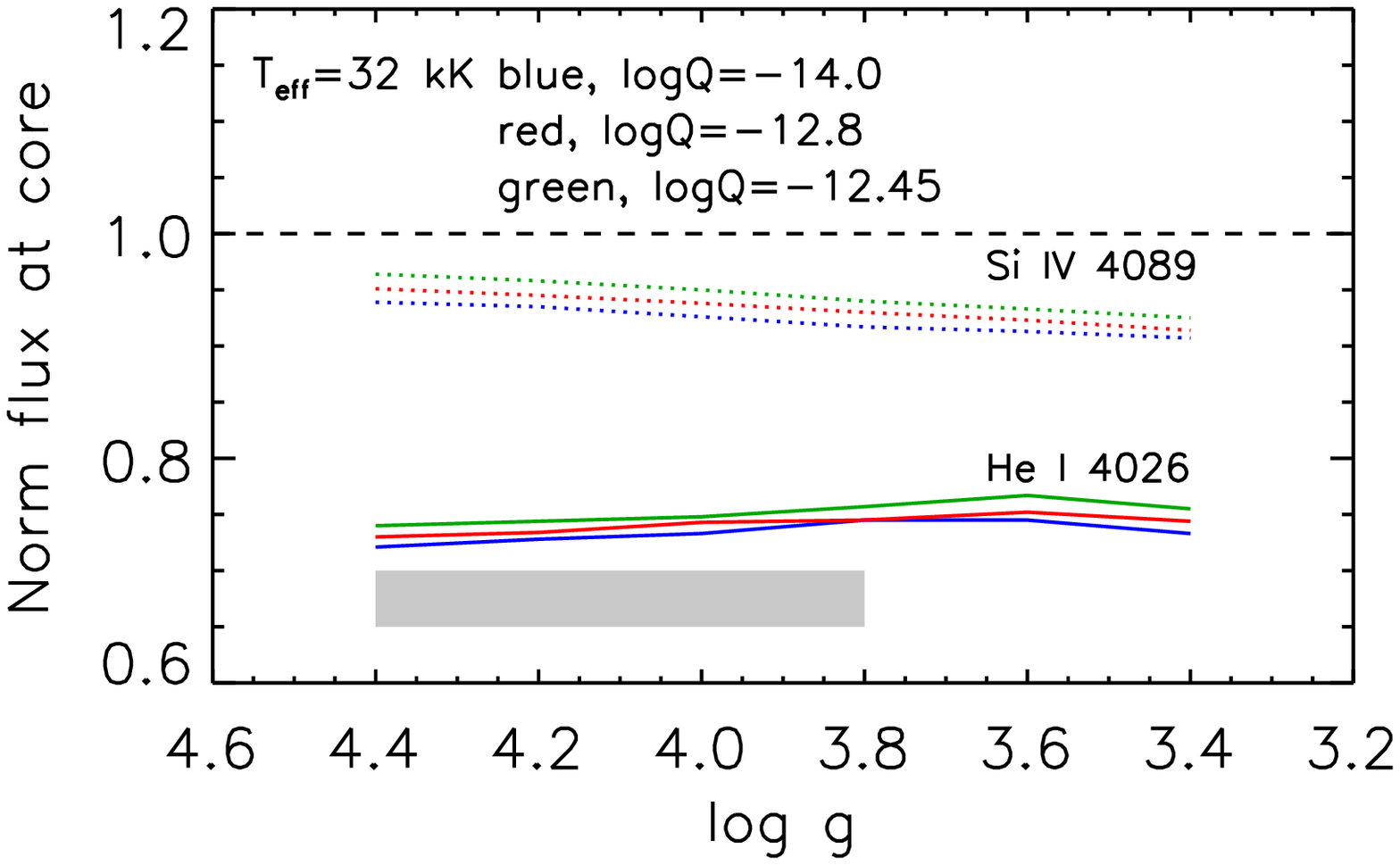}}
\caption{{\sc fastwind} model predictions for the line intensities
  used for luminosity criteria in late O-type stars as a function of
  surface gravity (\logg) for different \Teff-log\,$Q$ pairs and
  \epsSi\, = 7.0~dex.  Lines used in the primary and secondary
  criteria (see Table~\ref{LC}) are shown in the upper and lower
  panels, respectively. Line profiles were broadened using
  \vsini\,$=$\,100\,\kms, and a spectral resolving power of
  $R$\,$=$\,4\,000 (used to classify the VFTS spectra).  For all
  profiles, \vmic\ was fixed at 10 and 5~\kms for the NLTE and 
  formal integral computations, respectively. The \logg\ range for
  our sample is indicated by the grey rectangle.}
\label{fig6}
\end{figure*}

In general this was successful but there were a number of stars with a
notable discrepancy, and which ultimately led to the sample in the
current study. The \heiic/\heid\ ratio in the spectra of these stars
suggested a giant or bright giant luminosity class, but their \SiIVa\
lines were too weak for a giant and were more appropriate for a dwarf
(generally noted as `Si weak' by \citeauthor{walborn14}).

As noted earlier, the quality of the spectra may be an
issue\footnote{Stars flagged as Si weak by \citet{walborn14} comprised
  about 7\% of the AAA-rated O stars from the VFTS, and 18\% of the
  BBB-rated spectra. For the subsample analysed here, the
  corresponding fractions are 12\% and 88\%, respectively.}, but
\citet{walborn14} also suggested that the metallicity dependence of
the \SiIIIa/\heiib\ ratio for late O types, as well as the rapid
decline of the luminosity-dependent infilling of \heiic\ with
advancing type (such that \heiic/\heid\ has similar values near unity
at O9.7 II and B0 IV), might also contribute. Interestingly,
\citet{sota11} noticed a similar problem for some of their Galatic
objects and tentatively attributed it to unresolved binarity.

To maximise the usefulness of morphological classifications of hot 
massive stars, it is vital that we understand how they are influenced 
by  physical properties. Spectral type most  closely follows the stellar 
temperature, while luminosity class is more influenced by gravity 
(or, if \Teff is known, luminosity over the stellar mass).  
With this in mind we used the new HHeSi FW grids to investigate
the sensitivity of six of the diagnostic lines used in classification
of late O-type stars to changes in key stellar properties. In
particular, we measured the relative intensity at the core of each
line and followed its variation as a function of \Teff, \logg, and
log\,$Q$. The synthetic profiles were degraded to a resolving power
($R$) of 4\,000, as used to classify the VFTS targets, and convolved
with a rotational broadening profile corresponding to
\vsini\,$=$\,100\,\kms. Thus, the discussion below is strictly valid
for these specific values of \vsini\ and $R$ (for more information see
\citealt{markova11}) and for microturbulent velocities as discussed in
Sect.~\ref{para_determinations}. Models with \epsSi\,$=$\,7.0 and
7.2\,dex were considered.

\subsection{Predicted behaviour of luminosity criteria}
\label{lum_clsf}

Intensity predictions from the FW models for the line pairs used
as luminosity criteria in late O-type stars are shown in Fig.~\ref{fig6}. 
Over the parameter space of the models, the upper panels show that 
the depth of \heid\ does not react strongly to changes in \Teff, \logg,\ 
and log\,$Q$, while the intensity of \heiic\ is strongly sensitive to 
variations in these parameters. In particular, at \Teff\,$=$\,36\,kK 
(upper-left panel) the intensity of \heiic\ is (almost) exclusively 
sensitive to the wind strength, as suggested by \citet{sota11}; the 
stronger the wind, the weaker the line.  For these hotter temperatures 
and for the \logg\ values typical for unevolved  O-stars  
(marked by the grey rectangle), the ratio of \heiic\ to \heid\ is 
predicted to decrease monotonically from $\sim$2.8 for 
log\,$Q$\,$=$\,$-$14.0 to $\sim$2.1 for log\,$Q$\,$=$\,$-$12.8, and 
to about unity for log\,$Q$\,$=$\,$-$12.45, suggesting luminosity classes 
of V, III, and II, respectively \footnote{As the criteria listed in Tables~\ref{SpT}
  and \ref{LC} are qualitative but not quantitative, we performed a
  direct comparison to the standard stars from \citet{sota11} to
  connect the FW predictions with the corresponding spectral type or
  luminosity class.}.

For \Teff\,$<$\,36\,kK (middle and right-hand panels of Fig.~\ref{fig6}), 
in addition to the wind density the central depth of
\heiic\ is also sensitive to changes in \Teff\ and \logg, with the line
weakening at cooler temperatures. At \Teff\,$=$\,34\,kK, the
\heiic/\heid\ intensity ratio is still controlled by emission filling
in the He~{\scriptsize II} line and, as for the 36~kK models, it can
discriminate between luminosity class II (log\,$Q$\,$=$\,$-$12.45),
III (log\,$Q$\,$=$\,$-$12.85), and V (log\,$Q$\,$=$\,$-$14.0). In the
latter case, due to the effects of gravity, a luminosity
classification of III or IV (instead of V) is suggested for
4.2\,$\le$\,\logg$\,\le$\,4.4\,dex. Different behaviour is seen for
\Teff\,$<$\,34\,kK, where the depth of the \heiic\ line can be equal
to or slightly deeper than that of \heid, implying a luminosity
classification of II or III for all models, irrespective of their wind
strength or equatorial surface gravity. Interestingly,
\citet{martins18} also warned that the \heiic/\heid\ ratio is
basically the same for V, IV, and III luminosity classes in his
sample of late O-type (O9-O9.7) Galactic stars.

The lower panels of Fig.~\ref{fig6} show results for a similar
comparison of the predicted intensities for the lines in the secondary
luminosity diagnostic (i.e. the intensity ratio of \SiIVa\ to \heia).
The central depths of both lines are insensitive to changes in \Teff\
and log\,$Q$ (with variations of $<$3\%), but are (weakly) sensitive
to gravity.  In the models with \Teff\,$=$\,32\,kK the \SiIVa/\heia\
ratio decreases monotonically with gravity, suggesting a luminosity
class of IV or V for \logg\ values comparable to those of our targets
(and class III for \logg\,$\sim$\,3.4-3.6\,dex).

Two consequences of the above results at the cool edge of the O-type
temperature regime (\Teff\,$<$\,34\,kK) are as follows:
\begin{itemize}
\item[i)]{Use of the \heiic/\heid\ criterion can lead to a lack of
    class IV/V objects with \Teff\,$<$\,34\,kK, and the appearance of
    stars classified as class II/III stars but that have equatorial
    surface gravities that are typical for dwarfs (class V). For many
    of the latter, the so-called `Si~{\scriptsize IV} weak' phenomenon
    can emerge.}
\item[ii)]{At least for metallicity of the LMC, the intensity ratio of
    \SiIVa/\heia\ appears to provide more reliable luminosity classes
    (in the sense that the gravity of the star is traced more reliably 
    when combined with other information such as estimated absolute
    luminosities) compared to those originating from the \heiic/\heid\
    ratio.}
\end{itemize}

\begin{figure}
{\includegraphics[width=8.5cm,height=6.5cm]{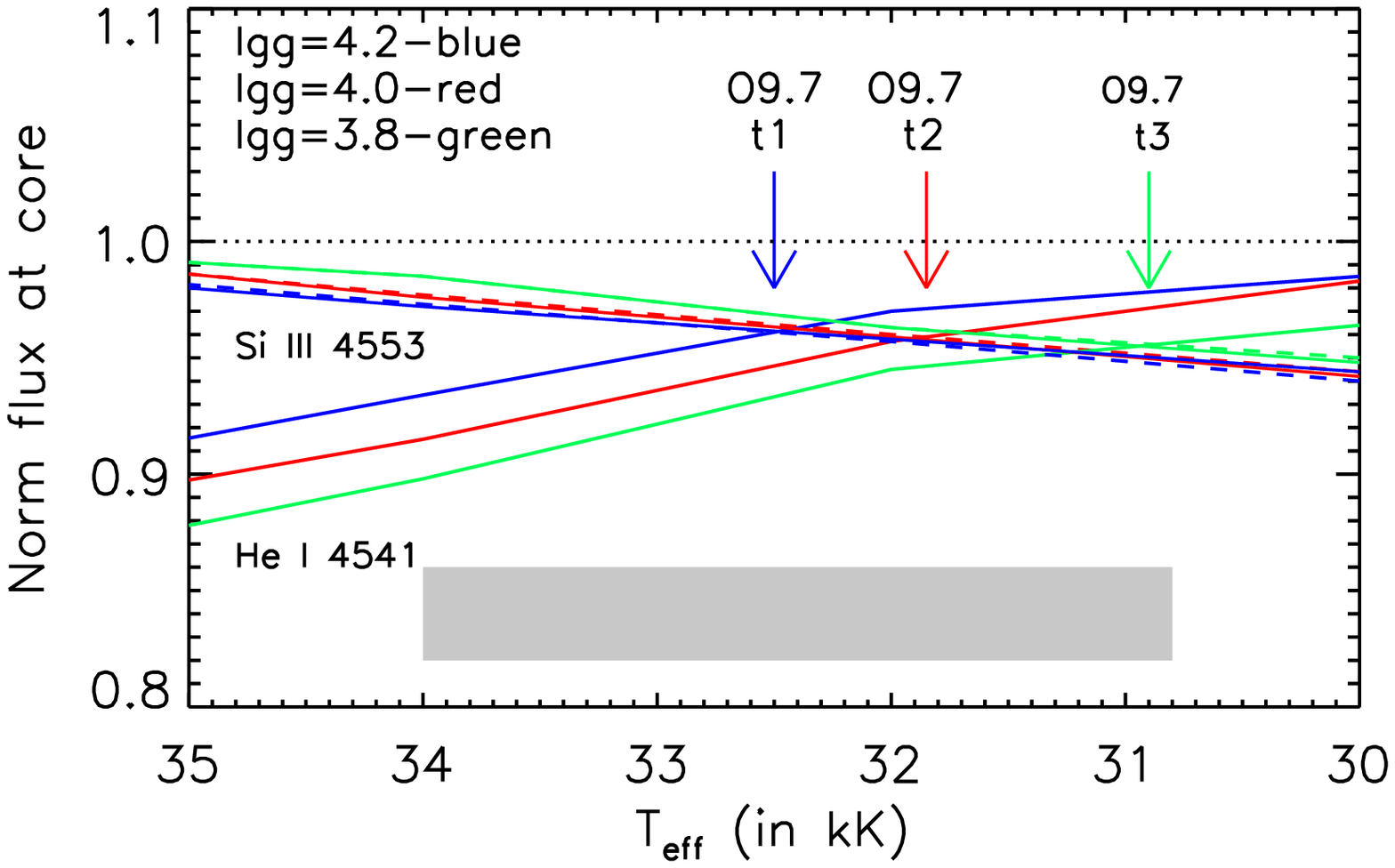}}
\caption{Predicted intensities for the lines involved in the spectral
  type criterion for late O stars in the LMC as a function of \Teff\
  for \epsSi\,$=$\,7.0\,dex and three values of \logg. The grey
  rectangle indicates the temperature range of our sample. The
  vertical arrows indicate the temperature at which the \SiIIIa/\heiib\
  ratio equals unity, which defines the O9.7 subtype at solar
  metallicity: t1\,$=$\,32.5\,kK, \logg\,$=$\,4.2; t2\,$=$\,31.85\,kK,
  \logg\,$=$\,4.0, and t3\,$=$\,30.9\,kK, \logg\,$=$\,3.8. The values of
  \vsini, $R$,  and \vmic\ are the same as in Fig.~\ref{fig6}.}
\label{fig7}
\end{figure}

 \subsection{Predicted behaviour of spectral type criteria}
\label{sp_clsf}

The predicted behaviour of the lines involved in the \SiIIIa/\heiib\
ratio, the classical spectral type diagnostic in the late O star
regime (see Table~\ref{SpT}), is shown in Fig.~\ref{fig7} as a
function of \Teff\ and \logg. Evidently, over the parameter space
covered by our sample, both lines are strongly temperature dependent
-- the central intensity of \SiIIIa\ strengthens towards cooler
\Teff\, with the opposite trend for the \heiib\ line. Additionally,
they are both sensitive to changes in gravity where the variability 
patters are again of an opposite nature to each other with  \SiIIIa\ being
less responsive than \heiib.

The main implication of these predictions is that for different
(\Teff, \logg) pairs, the same \SiIIIa/\heiib\ ratio can be observed.
For example, and as indicated in Fig.~\ref{fig7}, this ratio is equal 
to unity at: \Teff\,$=$\,32.5\,kK for \logg\,$=$\,4.2\,dex,
\Teff\,$=$\,31.85\,kK for \logg\,$=$\,4.0\,dex, and
\Teff\,$=$\,30.9\,kK for \logg\,$=$\,3.8\,dex. These temperatures fall
within the range of values derived for our sample (grey area in
Fig.~\ref{fig7}), so we conclude that the classification of our
targets as O9.5--O9.7 stars is compatible with their estimated \Teff\
and \logg\ values. Indeed, the \Teff--\logg\ results for our sample
(Fig.~\ref{fig8})  are in better agreement with those reported for 
other late O-type stars than with those found for early B-type stars.
 Their temperatures are in the range of those
derived for late O-type dwarfs in the LMC by \citet{gonzalez12} and
\citet{carolina17}, in contrast to the cooler temperatures (and
generally lower gravities) obtained for early B-type (B0-B2) dwarfs
from H07 and D18. Thus, it appears that, at least for our sample, the
morphological criteria used to assign spectral type (Table~\ref{SpT})
have not led to spurious results due to the lower metallicity of the
LMC.

\begin{figure}
{\includegraphics[width=8.5cm,height=6.2cm]{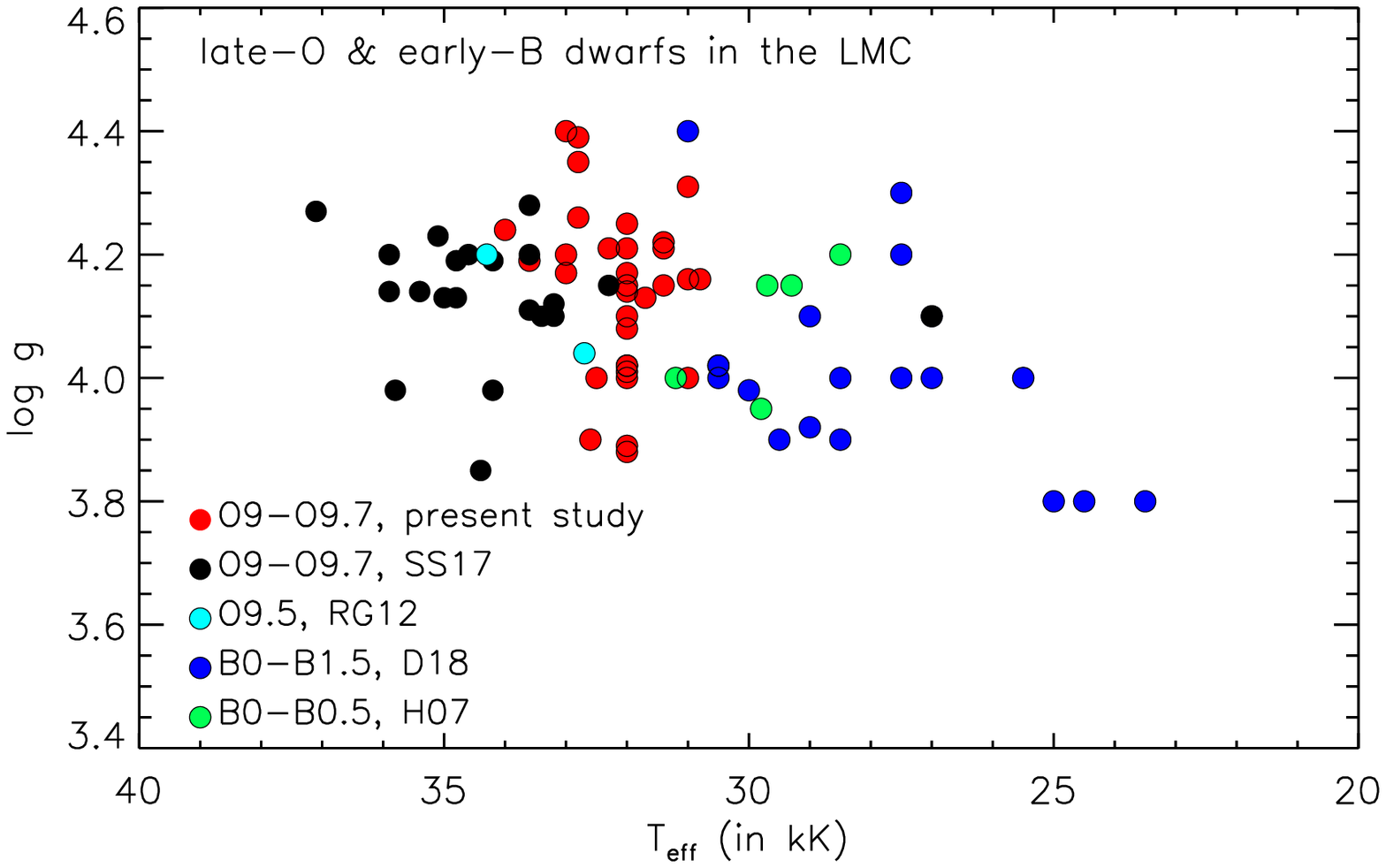}}
\caption{\Teff--\logg diagram of our sample stars compared to results
  in the LMC for early B-type dwarfs from H07 \citep{hunter07} and
  D18 \citep{dufton18} and for late O-type dwarfs from RG12
  \citep{gonzalez12} and SS17 \citep{carolina17}.}
\label{fig8}
\end{figure}

From Fig.~\ref{fig8} we also comment that our targets have comparable
gravities but are actually slightly cooler than the results from
\citet{carolina17}, with a maximum difference of 2 to 3\,kK for the
latest-type spectra (O9.7). Given that the same version of FW was
used in both analyses, we tentatively attribute these differences to
the adopted diagnostic lines (HHe vs. HHeSi) and fitting techniques
({\sc iacob-gbat} vs. by-eye fits).
Lastly, we note that the outcome of the predictions summarised above
do not change significantly if considering FW models with
\epsSi\,$=$\,7.2\,dex instead of 7.0\,dex.

\section{Summary}\label{conclusions}

Our motivation was to understand the properties of 32 late O-type
stars from the VFTS that were classified as giants or bright giants by
\citet{walborn14}, but for which the model-atmosphere analysis of RA17
found gravities that were more consistent with dwarf (or subdwarf)
stars. For many stars in the sample there was also a morphological
discrepancy between the \heiic/\heid\ and the \SiIVa/\heia\ line
ratios used to assign luminosity classes (with respect to Galactic
standards), which had led \citeauthor{walborn14} to flag them as
Si~weak.

Our analysis employed new HHeSi {\sc fastwind} models and a by-eye
fitting approach to revisit the physical properties of this intriguing
subset of late O-type stars. Our main conclusions can be summarised as
follows:

\begin{itemize}

\item{Our estimated physical parameters for the 32 stars are in good
    agreement with those from RA17 (aside from the He content). These
    results support the suggestion that these should be considered as
    low-luminosity, late O-type dwarfs (or subgiants) rather than
    giants or bright giants, as classified from their spectral
    morphology (i.e., using Galactic standards).}

\item{A reasonable fit to the helium lines was obtained for all stars
    using models with \Yhe\,$=$\,0.10\,$\pm$\,0.02 (and taking into
    account nebular contamination of He~I where relevant). This
    differs from RA17 where \Yhe\,$<$\,0.08 was derived for 
    $\sim$22\% of the sample (seven stars).  
    Differences in the adopted microturbulent velocities used in 
    both studies as well as weak or residual nebular contamination in 
    the He~I lines (which might not have been
    fully taken into account  during the automated fitting) may explain the previous 
    estimates (see Sect.~\ref{he_abn}).}

\item{Silicon abundances estimated for our sample range from 6.9 to
    7.2\,dex, with weighted means of 7.03\,$\pm$\,0.04 (NGC~2070),
    7.04\,$\pm$\,0.05 (NGC~2060) and 7.07\,$\pm$\,0.06\,dex (local
    field). These estimates are $\sim$0.1--0.2\,dex lower than those
    from {\sc tlusty} for B-type stars in two other clusters in the
    LMC, namely N~11 (H07) and NGC~2004 (T07). Aside from potential
    differences in atomic data and/or models, other contributing factors
    could be the different assumptions in the corresponding FW and {\sc
      tlusty} analyses, particularly regarding the microturbulent
    velocities (see Sects.  4.4.1 and 4.4.2), and/or real differences
    between the OB stars located in NGC~2060 and NGC~2070 in the
    30~Dor region, and those in N~11 and NGC~2004 (see Sect. 4.4.3).}
\end{itemize}

Prompted by these new results, we also investigated the predicted
behaviour of the spectral lines used to classify late O-type stars  
(Tables~\ref{SpT} and \ref{LC}), which resulted in the following findings.

\begin{itemize}
\item{At \Teff\,$<$\,34\,kK the strong dependence of the He~{\scriptsize II} 
    $\lambda$4686 line on \Teff\ and \logg\ (in addition to
    log\,$Q$) can lead to a \heiic/\heid\ ratio for a late O-star in
    the LMC that mimics a luminosity class of II or III, even for
    stars with high gravities and weak winds. Similar investigation of
    the \SiIVa/\heia\ ratio showed that it suggests luminosity
    class V for our targets, in better agreement with their physical
    properties from the model-atmosphere analysis.}

\item{The temperatures derived from our analysis are consistent with
    the expected values for stars classified as O9.5-O9.7.
    Additionally, their \Teff-- \logg\ distribution agrees with that
    of B-dwarfs (observed by both the VFTS and the FLAMES~I survey),
    in the sense that they constitute a sequence of objects with
    monotonically increasing temperatures and gravities.  This
    indicates that the \SiIIIa/\heiib\ ratio used to assign spectral
    type in Galactic stars can also provide reliable results for late
    O-type stars in the LMC (see Sect.~\ref{sp_clsf}).}
\end{itemize}

These tests of the spectral diagnostics have provided insights into
several puzzling issues revealed within the VFTS, namely: i) the
deficit of O-type dwarfs with \Teff\,$<$\,33\,kK \citep{carolina17};
ii) the appearance of late O-type giants or bright giants with surface
gravities more typical of dwarfs \citep{R17}, and iii) the discrepancy
between the primary and the secondary luminosity diagnostics for the
O9--O9.7 subtypes, leading to the `Si weak' stars from
\citet{walborn14}.

Finally,  although significantly improved, the
classification scheme proposed by \citet{sota11} is still qualitative
and is therefore prone to various complications. In particular, classification
of new stars requires a direct comparison with standards in order to assess the
strength of specific lines and line ratios used as diagnostics of
spectral type and luminosity class.  However, there are a number of
processes (e.g. mass loss, stellar rotation, microturbulence,
metallicity, and spectral resolving power) that can affect the
strength of those lines and line ratios (see e.g.  \citealt{markova09,
  markova11}), which can make classification of new stars difficult
(even in the Galaxy). To circumvent such complications,
\citet{martins18} recently quantified several classification criteria
based on archival spectra for 105 O-type Galactic stars.  A similar
study for O stars in the LMC would be highly valuable, and will be the
subject of a future study using the VFTS spectra.

\begin{acknowledgements}
  Based on observations at the European Southern Observatory Very
  Large Telescope in programme 182.D-0222. NM acknowledges financial
  support from the Bulgarian NSF (grant numbers DN08/1/13.12.2016 and
  DN 18/13/12.12.2017).
\end{acknowledgements}

\appendix
\section{Effects of microturbulent velocity on the diagnostic HHeSi lines}

As noted in Sect.~\ref{para_determinations}, the adopted microturbulence 
impacts the diagnostic HHeSi lines that were used to investigate our 
stellar sample at the cool edge of the O star domain. It affects both 
the NLTE occupation numbers (via a modified line-radiation field) and 
the emergent profiles via its direct impact on the formal integral.  
In our new models these two effects are treated separately. The two
examples shown in Figs.~\ref{A1} and \ref{A2} illustrate the impact of
the different values on the resulting line profiles.

\begin{figure*}
{\includegraphics[width=18cm]{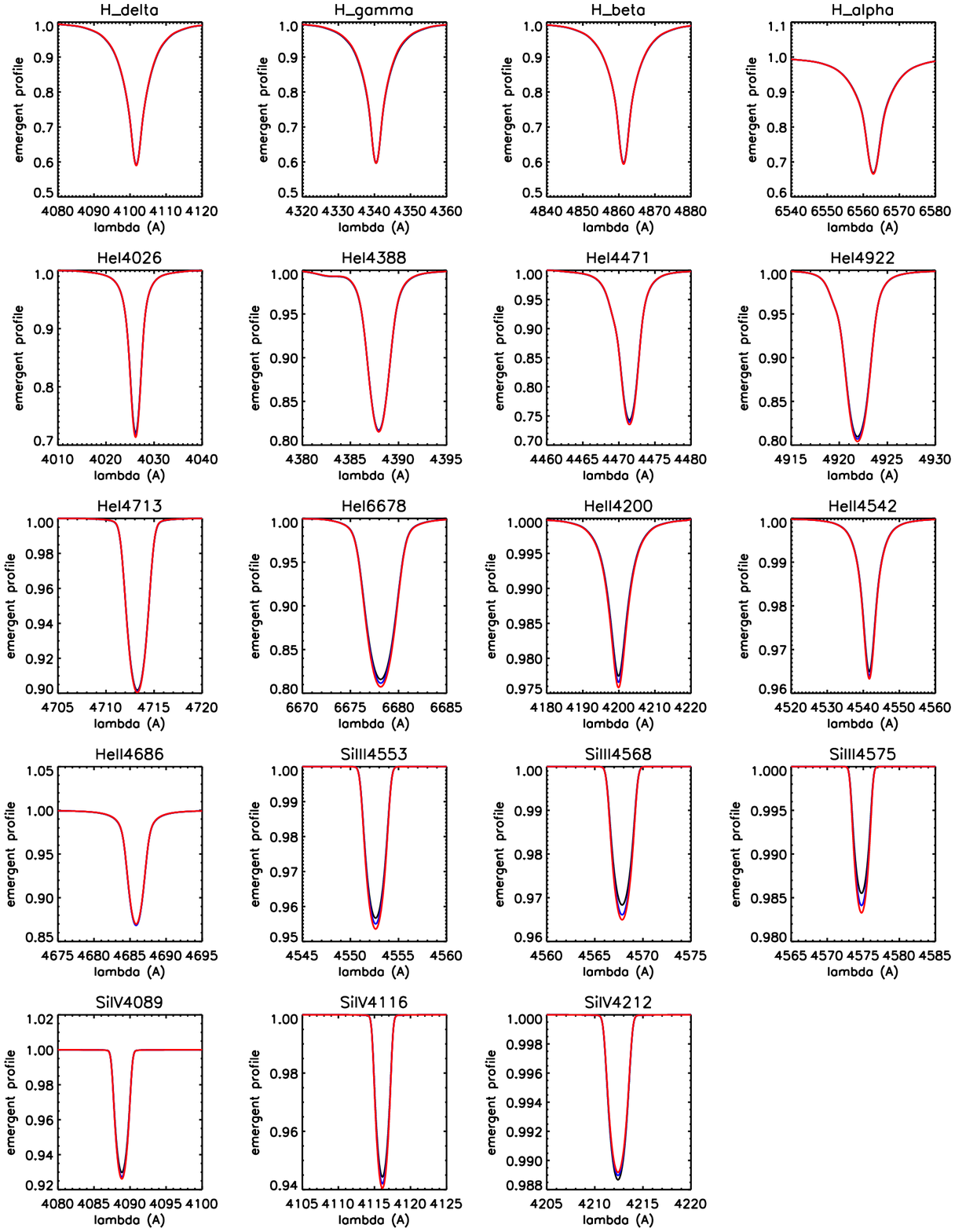}}
  \caption{Effect of microturbulence adopted in the NLTE calculations,
    \vmic(NLTE), on the strategic HHeSi lines for a model with typical
    parameters for our targets: \Teff\,$=$\,32\,kK,
    \logg\,$=$\,4.1\,dex, \Yhe\,$=$\,0.1, \epsSi=7.0\,dex, and log\,$Q$\,$=$\,$-$13.5.
    The profiles were computed assuming \vmic(NLTE)\,=\,5~\kms
    (black), 10~\kms (blue) and 15~\kms (red).  All profiles have been
    degraded to the resolving power of the VFTS Medusa spectra
    ($R$\,$\sim$8\,000) and convolved with a rotational profile
    corresponding to \vsini\,$=$\,100\,\kms. In order to isolate
    statistical equilibrium effects, a fixed value of
    \vmic\,$=$\,5\,\kms for the formal integral was used in these
    examples.}\label{A1}
\end{figure*}

\begin{figure*}
{\includegraphics[width=18cm]{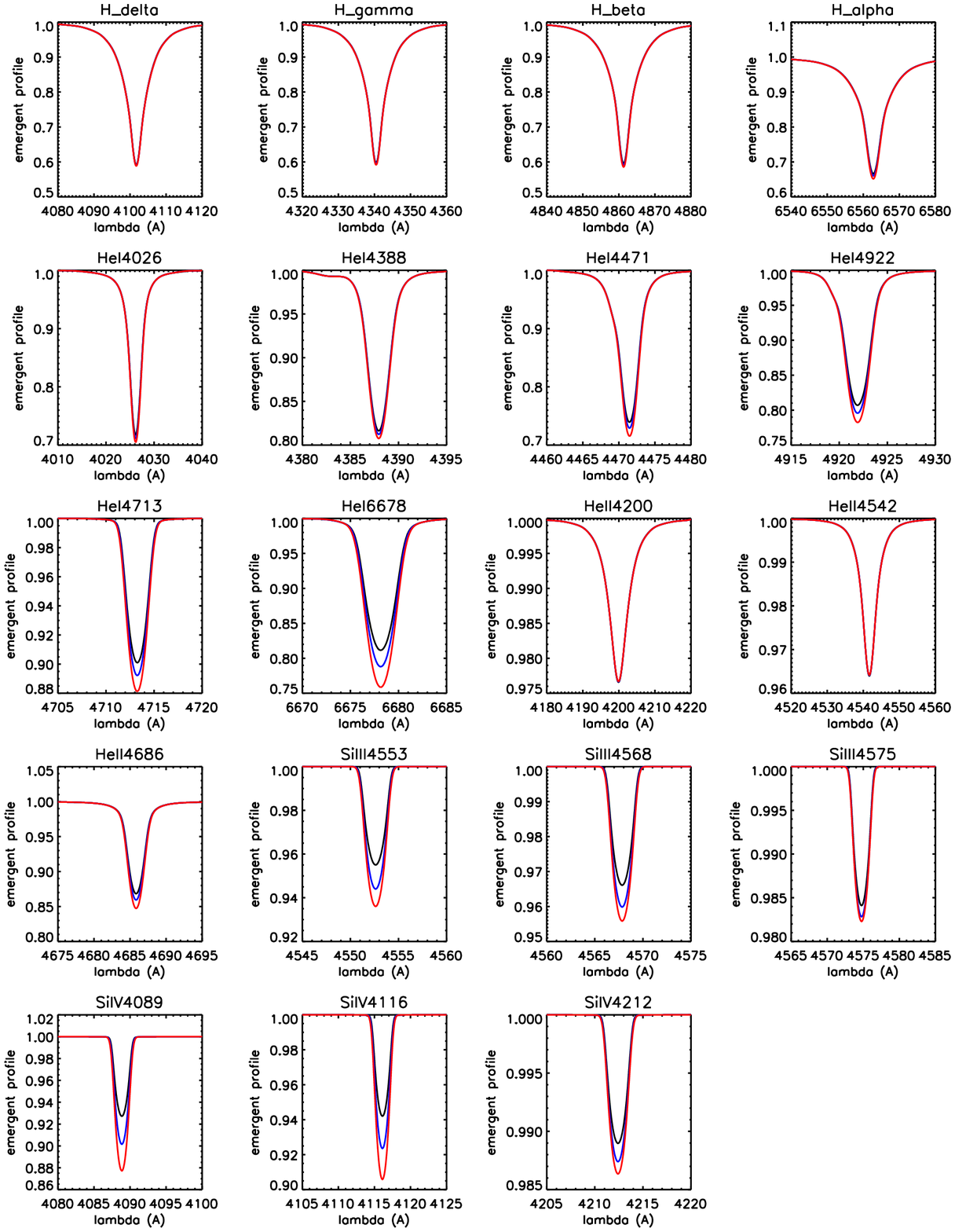}}
\caption{Effect of microturbulence adopted in the formal integral 
  computations on strategic HHeSi lines for a model  with typical 
  parameters for our targets: \Teff\,$=$\,32\,kK, \logg\,$=$\,4.1\,dex, 
  \Yhe\,$=$\,0.1, \epsSi=7.0\,dex and log\,$Q$\,$=$\,$-$13.5. The profiles 
  were computed 
  for \vmic\,=\,5\,\kms (black), 10\,\kms (blue), 15~\kms(red). All 
  profiles have been degraded to the resolving power of 
  the VFTS Medusa spectra ($R$\,$\sim$8\,000) and convolved with a rotational 
  profile corresponding to \vsini\,$=$\,100\,\kms. In order to isolate radiative 
  transfer effects, a fixed value of \vmic\,$=$\,10\,\kms for the NLTE occupation 
  numbers  was consistently adopted in these examples.}
\label{A2}
\end{figure*}

\end{document}